\documentclass[11pt]{article}

\usepackage{tikz}
\usetikzlibrary{arrows}
\usepackage{epsfig,subfigure}
\usepackage{amssymb}
\usepackage[fleqn]{amsmath}
\usepackage{amsthm}
\usepackage{color}
\newcommand{\define}{\stackrel{\triangle}{=}}
\def\QED{\mbox{\rule[0pt]{1.5ex}{1.5ex}}}

\usepackage{graphicx}

\usepackage{amssymb}
\usepackage{amsmath,amsfonts,amssymb}
\usepackage{verbatim}
\usepackage{stfloats}
\usepackage[bookmarks=false]{}

\newtheorem{theorem}{Theorem}
\newtheorem{corollary}{Corollary}[theorem]
\newtheorem{definition}{Definition}
\newtheorem{lemma}{Lemma}

\newtheorem{observation}{Observation}
\newtheorem{example}{Example}
\newcommand\blfootnote[1]{%
  \begingroup
  \renewcommand\thefootnote{}\footnote{#1}%
  \addtocounter{footnote}{-1}%
  \endgroup
}

\begin{document}
\date{}
\title{%Blind Interference Alignment for \\ Private Information Retrieval
On the Capacity of Locally Decodable Codes
%\thanks{This work is supported by NSF grants CCF-1317351 and CCF-0963925.}
}
%\author{\IEEEauthorblockN{Hua Sun and Syed A. Jafar}
%\IEEEauthorblockA{Center for Pervasive Communications and Computing (CPCC)\\
%University of California Irvine, Irvine, CA 92697
%%\\ Email: \{huas2, syed\}@uci.edu
%}}
\author{ %\IEEEauthorblockN{Hua Sun and Syed A. Jafar}
%\IEEEauthorblockA{Center for Pervasive Communications and Computing (CPCC)\\
%University of California Irvine, Irvine, CA 92697
\normalsize Hua Sun and Syed A. Jafar \\
%{\small Center for Pervasive Communications and Computing (CPCC)}\\
%{\small University of California Irvine, Irvine, CA 92697}\\
%{\small \it Email: \{huas2, syed\}@uci.edu}
}
%}

\maketitle

\blfootnote{Hua Sun (email: hua.sun@unt.edu) is with the Department of Electrical Engineering at the University of North Texas. Syed A. Jafar (email: syed@uci.edu) is with the Center of Pervasive Communications and Computing (CPCC) in the Department of Electrical Engineering and Computer Science (EECS) at the University of California Irvine. }

\begin{abstract}
A locally decodable code (LDC) maps $K$ source symbols, each of size $L_w$ bits, to $M$ coded symbols, each of size $L_x$ bits, such that each source symbol can be decoded from $N \leq M$ coded symbols. A perfectly smooth LDC further requires that each coded symbol is uniformly accessed when we decode any one of the messages. The ratio $L_w/L_x$ is called the symbol rate of an LDC. The highest possible symbol rate for a class of LDCs is called the capacity of that class. It is shown that given $K, N$, the maximum value of capacity of perfectly smooth LDCs, maximized over all code lengths $M$, is $C^*=N\left(1+1/N+1/N^2+\cdots+1/N^{K-1}\right)^{-1}$. 
Furthermore, given $K, N$, the minimum code length $M$ for which the capacity of a perfectly smooth LDC is $C^*$ is shown to be $M = N^K$. Both of these results generalize to a broader class of LDCs, called universal LDCs. The results are then translated into the context of PIR$_{\max}$, i.e., Private Information Retrieval  subject to maximum (rather than average) download cost metric. It is shown that the minimum upload cost of capacity achieving PIR$_{\max}$ schemes is $(K-1)\log N$. The results also generalize to a variation of the PIR problem, known as Repudiative Information Retrieval (RIR).\end{abstract}

\newpage
\allowdisplaybreaks
\section{Introduction}
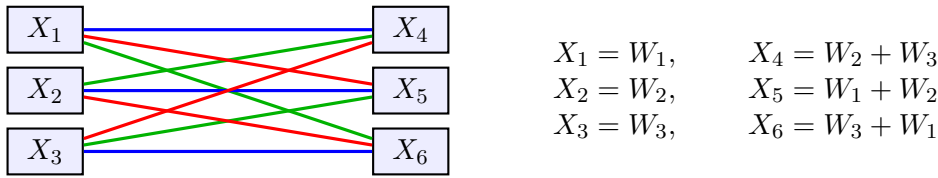
\begin{figure}[!h]
\begin{center}
\noindent\begin{tikzpicture}[scale=0.81]
\node (X1) at (0,3) [draw,thick,minimum width=1cm,minimum height=0.61cm, fill=blue!7] { $X_1$};
\node (X2) at (0,2) [draw,thick,minimum width=1cm,minimum height=0.61cm, fill=blue!7] {$X_2$};
\node (X3) at (0,1) [draw,thick,minimum width=1cm,minimum height=0.61cm, fill=blue!7] {$X_3$};
\node (X4) at (6,3) [draw,thick,minimum width=1cm,minimum height=0.61cm, fill=blue!7] {$X_4$};
\node (X5) at (6,2) [draw,thick,minimum width=1cm,minimum height=0.61cm, fill=blue!7] {$X_5$};
\node (X6) at (6,1) [draw,thick,minimum width=1cm,minimum height=0.61cm, fill=blue!7] { $X_6$};
\draw[very thick, black!30!green] (X1)--(X6);
\draw[very thick, black!30!green] (X2)--(X4);
\draw[very thick, black!30!green] (X3)--(X5);
\draw[very thick, blue] (X1)--(X4);
\draw[very thick, blue] (X2)--(X5);
\draw[very thick, blue] (X3)--(X6);
\draw[very thick, red] (X1)--(X5);
\draw[very thick, red] (X2)--(X6);
\draw[very thick, red] (X3)--(X4);
\node (def) at (11.5,2){$\begin{array}{lll}
X_1=W_1,&~~&X_4=W_2+W_3\\
X_2=W_2,&&X_5=W_1+W_2\\
X_3=W_3,&&X_6=W_3+W_1
\end{array}
$};
\end{tikzpicture}
\end{center}
\caption{\small \it An SLDC with locality $N=2$ that encodes $K=3$ binary ($L_w=1$) source symbols, $W_1, W_2, W_3$, into $M=6$ binary ($L_x=1$) coded symbols, $X_1, \cdots, X_6$. }\label{fig:sldc}
\end{figure}

A locally decodable code (LDC) with locality $N$ is a mapping from $K$ source symbols, $\mathcal{W}=\{W_1, W_2, \cdots, W_K\}$, each of size $L_w$ bits, to $M$ coded symbols, $\mathcal{X}=\{X_1, X_2, \cdots, X_M\}$, each of size $L_x$ bits, such that for every source symbol $W_k$, there exists at least one subset of $N$ coded symbols, $S\subset \mathcal{X}$, $|S|=N$, such that $W_k$ can be recovered from the elements of $S$. Such a set $S$ is called a decoding set for $W_k$. This basic definition is somewhat trivial, for example, any systematic code is locally decodable with locality $N=1$. LDCs are useful primarily if they are capable of withstanding a significant fraction of corrupted coded symbols without losing their local decodability. An $(N, \delta, \epsilon)$ LDC is guaranteed to have locality $N$ and a randomized decoding algorithm that succeeds with probability at least $1-\epsilon$ when the fraction of corrupted coded symbols is at most $\delta$. For this  to be meaningful, there must be multiple decoding sets for each source symbol. Let $\mathcal{S}_k$ be the set of decoding sets for source symbol $W_k$, so that if $S\in\mathcal{S}_k$ then $S\subset \mathcal{X}$, $|S|=N$, and $W_k$ is decodable from $S$. An LDC is said to be \emph{perfectly smooth} if the coded symbols are uniformly distributed across decoding sets. Specifically, $\forall m_1,m_2\in\{1,2,\cdots, M\}$, and $\forall k\in\{1,2,\cdots, K\}$, the number of decoding sets in $\mathcal{S}_k$ that contain $X_{m_1}$, must be equal to the number of decoding sets in $\mathcal{S}_k$ that contain $X_{m_2}$. If there are $|\mathcal{S}_k|$ decoding sets for $W_k$ in a perfectly smooth LDC (SLDC) with locality $N$, then every coded symbol must appear in exactly $N|\mathcal{S}_k|/M$ of them. For such a code, at least one uncorrupted decoding set survives as long as the fraction of corrupted coded symbols, $\delta$, is less than $1/N$. This is because each corrupted coded symbol can corrupt at most $N|\mathcal{S}_k|/M$ decoding sets in $\mathcal{S}_k$. If $\delta M$ coded symbols are corrupted, then the number of decoding sets that are corrupted is no more than $\delta N |\mathcal{S}_k|$. So a decoding algorithm that randomly chooses one of the decoding sets must be successful with probability at least $1-\delta N$,  provided that $\delta<1/N$. Therefore, an SLDC is  an $(N, \delta, 1-\delta N)$ LDC for any $\delta<1/N$.  By the same token, the minimum distance $d$ of an SLDC, i.e., the minimum number of coded symbols that must be erased for a loss of data to occur, is at least $M/N$. Figure \ref{fig:sldc} shows an example of an SLDC  with locality $N=2$ that encodes $K=3$ binary ($L_w=1$) source symbols, $W_1, W_2, W_3$, into $M=6$ binary ($L_x=1$) coded symbols, $X_1, \cdots, X_6$.
 The decoding sets for $W_1, W_2, W_3$ are comprised of pairs of coded symbols connected by blue, red, and green edges, respectively. This is  also a $(2, \delta, 1-2\delta)$ LDC for  $\delta<1/2$. So if $\delta=1/3$, and any two coded symbols $X_i,X_j$ are corrupted, then at least one of the three decoding sets remains uncorrupted for every source symbol, and a randomized decoder succeeds with probability at least $1-\delta N=1/3$. The minimum distance of this code is  $d=M/N=3$ because, e.g., a loss of $X_1, X_5, X_6$ causes a loss of data ($W_1$ is lost).

LDCs were introduced in the year 2000 by Katz and Trevisan in  \cite{LDC}. One of the motivations for studying LDCs comes from distributed storage applications. Coding is used in distributed storage systems to limit storage and decoding costs  while providing resilience against failures of storage nodes and efficient repair when such failures occur. LDCs are especially effective for reducing the decoding cost in commonly encountered scenarios where multiple datasets are jointly encoded and only one of them needs to be retrieved. In particular, smoothness of LDCs is a desirable feature for distributed storage because it minimizes risk by spreading it evenly across storage nodes. Remarkably,  LDCs play  even more important roles in  complexity theory \cite{Trevisan}, \cite[Chapters~17, 18]{Arora_Barak}, data structures \cite{DeWolf, Gal_Mills},  fault tolerant computation \cite{Romashchenko}, multiparty computation \cite{Ishai_Kushilevitz} and private information retrieval (PIR) \cite{PIRfirst, Beimel_Ishai_Kushilevitz, YekhaninPhd}. As such, understanding the fundamental limits of LDCs (especially the tradeoff between code length $M$ and locality $N$) is recognized as a major open problem in theoretical computer science  \cite{Ishai_Kushilevitz}, whose answer could have a domino effect on a number of related problems.
For further details on LDCs, we refer to the excellent tutorials in \cite{Yekhanin_FnT, Kopparty_Saraf} and references therein.

In this work we view this open problem through the lens of PIR. In its basic form \cite{PIRfirst}, PIR is the problem of efficiently retrieving a desired message from a set of  $K$ messages that are replicated across $N$ non-colluding databases, without disclosing any information about the identity of the desired message to any individual database. The strong connection between PIR and LDCs is evident from the example illustrated in Figure \ref{fig:sldc}.  
 In fact the example is derived from a PIR scheme with $K=3$ messages, $W_1, W_2, W_3$, and two databases that store $(X_1, X_2, X_3)$ and $(X_4, X_5, X_6)$, respectively. The user randomly asks Database $1$ for one of $X_1, X_2$ or $X_3$, and  asks Database $2$ for the other element of the decoding set for his desired message, which is also uniformly distributed over $X_4, X_5, X_6$, thus revealing no information to either database about which message is being retrieved. The upload cost for this PIR scheme is a $3$-ary symbol per database. Interestingly, as shown in \cite{Sun_Jafar_PIR},  the capacity of PIR subject to this upload cost is $1/2$, so the scheme shown in Figure \ref{fig:sldc} is optimal among all PIR schemes with the same upload constraint. %Remarkably, all PIR schemes produce smooth locally decodable codes.

In particular, this work is motivated by recent capacity characterizations of PIR with various assumptions on message sets, storage, and upload costs \cite{Sun_Jafar_PIR, Tian_Sun_Chen_Upload, Sun_Jafar_PIRL, Sun_Jafar_TPIR, Tajeddine_Gnilke_Karpuk_Etal, Tajeddine_Rouayheb, Banawan_Ulukus}. The capacity of PIR, $C_{\mbox{\tiny PIR}}(N, K)$, is the maximum number of bits of desired message that can be retrieved per bit of total download from the $N$ databases. Defining $R_s=L_w/L_x$ as the \emph{symbol rate}  of an LDC,  the corresponding notion of capacity, $C_{\mbox{\tiny LDC}}(M, N, K)$,  is the maximum symbol rate that is feasible for an LDC given the locality parameter $N$, the code length $M$, and the number of source symbols $K$. From this perspective, the fundamental tradeoff for SLDCs is expressed in terms of the $4$ parameters: $M, N, K, R_s$.  It is desirable for $M,N$ to take smaller values, and for $K, R_s$ to take larger values. The rate $R_s$ is a critical part of this tradeoff. If we consider $M, K$  as independently chosen natural numbers, then the range of values of $N$ is between $1$ and $M$, while the range of values of $R_s$ is between $1/K$ and $M/K$. At one extreme, $N=1$ forces $R_s = 1/K$. This is because $N=1$ for an SLDC implies that \emph{all} source symbols can be decoded from any single coded symbol. At the other extreme, $R_s= M/K$ forces $N=M$, because there is no redundancy, i.e., the total number of bits of all coded symbols is the same as the total number of bits of all source symbols. 

In this paper we explore two particular aspects of the $(M, N, K, R_s)$ tradeoff. The first is  the tradeoff between $N, K, R_s$ for  unconstrained $M$. In other words, we identify the capacity of an SLDC for arbitrary $N, K$ and unconstrained code length $M$. Specifically we show that,
\begin{align}
C^*(N,K)\triangleq \max_{M\in\mathbb{N}} C_{\mbox{\tiny LDC}}(M, N, K)&=N\left(1+\frac{1}{N}+\cdots\frac{1}{N^2}+\cdots+\frac{1}{N^{K-1}}\right)^{-1}
\end{align} The second aspect of the tradeoff that we characterize is  the minimum codeword length $M^*$ that is needed to achieve $C^\star(N, K)$ for arbitrary $N, K$. Specifically, we show that $M^*=N^K$. Remarkably, both results are shown not only for all SLDCs but also for a broader class of LDCs that we label \emph{universal} LDCs (ULDCs). An LDC is \emph{universal} if every coded symbol appears in at least one of the decoding sets of every source symbol. Mathematically, a ULDC is defined by the property that $\forall m\in\{1,2,\cdots,M\}$, and $\forall k\in\{1,2.\cdots, K\}$, there exists some $S\in\mathcal{S}_k$ such that $X_m\in S$. Clearly, every SLDC is a ULDC. However,  not every ULDC is an SLDC. For example, the LDC that maps $K=3$ binary source symbols $W_1, W_2, W_3$ to  the $M=4$ binary code symbols $W_1, W_2, W_3, W_2+W_3$ with locality $N=2$ and decoding sets $\mathcal{S}_1=\{\{W_1, W_2\}, \{W_1, W_3\}, \{W_1, W_2+W_3\} \}$, $\mathcal{S}_2=\{\{W_1, W_2\}, \{W_2, W_3\}, \{W_3, W_2+W_3\} \}$ and $\mathcal{S}_3=\{\{W_1, W_3\}, \{W_2, W_3\}, \{W_2, W_2+W_3\} \}$, is universal but not perfectly smooth. While less structured than SLDCs, evidently ULDCs retain all the structure needed for the two aspects of the tradeoff that are explored in this work.

For our final result, we apply the new insights from the study of fundamental limits of LDCs back to the problem of PIR. Recall that the rate of a PIR scheme is defined as $R_p=\frac{L_w}{ND}$, where $L_w$ is the number of bits of each message, $N$ is the number of databases, and $D$ is the number of bits downloaded from each database. For most PIR capacity results \cite{Sun_Jafar_PIR, Sun_Jafar_TPIR, Banawan_Ulukus, Sun_Jafar_SPIR} the parameter $D$ may be interpreted either as the \emph{average} download per database or as the \emph{maximum} download from any database (maximized across all databases and all queries), without changing the capacity. This is because the normalized downloads for almost all PIR schemes are either already identical across databases or can be made identical by time-sharing across different permutations of databases. Exceptions include  \cite{Sun_Jafar_PIRL} which admits only the maximum download formulation and  \cite{Tian_Sun_Chen_Upload} which allows only the average download formulation.  Reference \cite{Sun_Jafar_PIRL} considers the capacity of PIR for fixed length messages, and relies on the  maximum download formulation because averages are less meaningful over the finite horizon. Reference \cite{Tian_Sun_Chen_Upload} on the other hand considers the minimum upload cost of a capacity achieving PIR scheme, and allows only the average download formulation because the PIR scheme is asymmetric and the usual approach of making the scheme symmetric with  time-sharing arguments does not work (does not preserve the upload cost).  When PIR is viewed in relation to LDCs, the natural interpretation of $D$ is the maximum download across all servers and all queries,\footnote{Equivalently, the size of the download from each database $n$ is fixed at the same constant value, $D$, for all queries and all databases, $n\in\{1,2,\cdots,N\}$.} which corresponds to $L_x$ in the corresponding LDC setting.  To make the distinction clear, we refer to PIR with the maximum download metric as PIR$_{\max}$, and PIR with the average download metric as PIR$_{\mbox{\footnotesize ave}}$. 
Using insights from LDCs, we determine the minimum upload cost needed to achieve the capacity of PIR$_{\max}$. Specifically, we show that  the minimum upload  for \emph{any} capacity achieving PIR$_{\max}$ scheme, linear or non-linear,  is $(K-1)\log N$ bits per database, i.e., the user must upload a $q$-ary symbol per database where $q$ is at least $N^{K-1}$.  Our result complements  the result of \cite{Tian_Sun_Chen_Upload} which shows that  the minimum upload cost for capacity achieving PIR$_{\mbox{\footnotesize ave}}$ schemes is also $(K-1)\log N$ bits per database, although the optimality in \cite{Tian_Sun_Chen_Upload} is established only  within a restricted class of decomposable (e.g., linear) schemes. Remarkably, while the capacity and minimum upload cost characterizations are identical for PIR$_{\max}$ and PIR$_{\mbox{\footnotesize ave}}$, the mapping between the corresponding PIR schemes turns out to be highly non-trivial. Furthermore, just as our results for SLDCs generalize to ULDCs, by the same token we show that both the capacity and  the minimum upload cost are unaffected if the privacy constraint is relaxed in the PIR$_{\max}$ problem formulation from perfect privacy to a weaker \emph{deniability} condition. Perfect privacy implies that the query to each database must not reveal any information about the user's desired message index.  Deniability only implies that the query does not absolutely rule out any message from being the user's desired message, i.e., even if some messages are revealed by the query to be more likely to be the desired message than others, each message has a non-zero probability of being the desired message. Information retrieval under a deniability constraint is called \emph{Repudiative} information retrieval (RIR) in \cite{RIR}. Surprisingly, under the maximum download formulation, PIR$_{\max}$ and RIR$_{\max}$ have the same\footnote{Under the average download formulation, the capacity of PIR$_{\mbox{\tiny ave}}$ is not the same as the capacity of RIR$_{\mbox{\tiny ave}}$. In particular, the  capacity of RIR$_{\mbox{\tiny ave}}$ is trivially seen to be  $1$ if the number of databases is $N>1$. For example, let $(i,j)$ be a random permutation of $(1,2)$ generated privately by the user. The user downloads his desired message $W_\theta$ from Database $i$. With probability $\epsilon$ the user downloads a randomly chosen undesired message $W_{\theta'}$ from Database $j$. It is easy to verify that the scheme is valid for RIR, and that the rate achieved under the average download formulation with this scheme is $1/(1+\epsilon)$ which approaches $1$ as $\epsilon\rightarrow 0$. If $N=1$ then the capacity of RIR is $1/K$, same as PIR, under both average and maximum download formulations.} capacity, and the same minimum upload cost.

{\it Notation: For positive integers $n_1, n_2$, with $n_1 \leq n_2$, we use the notation $[n_1:n_2]$ to represent the set $\{n_1,n_1+1,\cdots, n_2\}$. For a set $A$, $|A|$ denotes its cardinality and $X_{A}$ represents the set $\{X_{i}, i \in A\}$. For two random variables $X, Y$, the notation $X \sim Y$ denotes that $X$ and $Y$ are identically distributed.} 

\section{Problem Statement and Preliminaries}\label{sec:model}
\subsection{Locally Decodable Codes (LDC)}
\begin{definition}[Set of Source Symbols, $\mathcal{W}$] Define $\mathcal{W}=\{W_1, \cdots, W_K\}$ as a set of $K$ independent source symbols, each of size $L_w$ bits, 
\begin{align}
& H(W_1, \cdots, W_K) = H(W_1) + \cdots + H(W_K), \label{h1}\\
& L_w= H(W_1) = \cdots = H(W_K). \label{h2}
\end{align}
\end{definition}
\begin{definition}[Set of Coded Symbols, $\mathcal{X}$] Define $\mathcal{X}=\{X_1, X_2, \cdots, X_M\}$ as a set of $M$ coded symbols each of size $L_x$ bits,
\begin{align}
L_x&=H(X_1) = \cdots = H(X_M). \label{xx}
\end{align}
\end{definition}
Note that $L_x$ and $L_w$ are not necessarily integer values. For example, if $W_i$ are uniformly random $3$-ary symbols, then $L_w=\log(3)$ bits. Furthermore, both $L_w$ and $L_x$ are allowed to take arbitrarily large values, since it is only their relative size that matters (see Definition 6). Indeed, in typical applications, such as distributed storage, each source symbol may represent a large dataset and each coded symbol may represent all data stored in one storage node. Measuring the size of each symbol by its entropy is especially meaningful  for large symbols which can be optimally compressed.

\begin{definition}[LDC $(\mathcal{C}, \mathcal{S}_{[1:K]})$] \label{def:3}
An LDC $(\mathcal{C}, \mathcal{S}_{[1:K]})$ with locality $N$ is comprised of a mapping $\mathcal{C}$ from $(W_1, \cdots, W_K)$ to  $(X_1, \cdots, X_M)$,  and $K$ non-empty sets $\mathcal{S}_k, k\in[1:K]$, called decoding supersets. Elements of the decoding superset $\mathcal{S}_k$ are called decoding sets of the source symbol $W_k$. Each decoding set of $W_k$ is itself a set $S$ containing $N$ coded symbols from which $W_k$ can be recovered.
\begin{align}
S\in\mathcal{S}_k&\Rightarrow 
\left\{
\begin{array}{rl}
S&\subset\mathcal{X},\\
|S|&=N,\\ 
H(W_k\mid S)&=0.
\end{array}
\right.
\end{align}
\end{definition}
Definition \ref{def:3} is useful only as a baseline upon which the definitions of more interesting types of LDCs can be built. The most interesting type of LDCs for our purpose are perfectly smooth LDCs, defined next.

\begin{definition}[Perfectly Smooth LDC (SLDC)] \label{def:4} An LDC is said to be perfectly smooth if for all $k\in[1:K]$, a uniform choice of a decoding set from  $\mathcal{S}_k$  implies that each coded symbol is equally likely to be in the chosen decoding set. Equivalently,  $\forall m, m'\in[1:M]$ and $\forall k\in[1:K]$,
\begin{align}
\left|\{S \mid S\in \mathcal{S}_k, X_m\in S\} \right|&=\left|\{S \mid S\in \mathcal{S}_k, X_{m'}\in S\} \right|
\end{align}
\end{definition}
Thus, in an SLDC, every coded symbol appears in the same number of decoding sets for any given source symbol. While SLDCs are most commonly encountered in various applications of LDCs, it is useful to also define a broader class of LDCs, called universal LDCs.
\begin{definition}[Universal LDC (ULDC)] An LDC is said to be universal if every coded symbol $X_m, m\in[1:M]$ appears in at least one of the decoding sets of every source symbol $W_k, k\in[1:K]$.
\begin{eqnarray}
\forall m \in [1:M], ~\forall k \in [1:K], ~\exists S\in \mathcal{S}_{k} ~\mbox{such that}~X_m \in S.
\end{eqnarray}
\end{definition}
Note that an SLDC is universal by definition.
\begin{definition}[Symbol Rate and Capacity] The symbol rate of an LDC is defined as,
\begin{align}
R_s &=  \frac{L_w}{L_x}, \label{eq:rldc}
\end{align}
and the supremum of $R_s$ values achievable within a class of LDCs is called the capacity of that class of LDCs.
\end{definition}
For example, it may be of interest to find the capacity of the class of SLDCs for given values of locality parameter $N$, the number of source symbols $K$, and the code length $M$. Another important quantity of interest is the code rate of an LDC, 
\begin{align}
R_c &= \frac{KL_w}{ML_x}
\end{align}
which measures the redundancy of the code. Note that $R_c = \frac{K}{M}R_s$.

\subsection{Private Information Retrieval (PIR$_{\max}$)}
Instead of repeating the definition of the PIR problem from, say \cite{Sun_Jafar_PIR}, let us present it through the following definitions that parallel the corresponding notions in the context of LDCs. As much as possible we will use the same notation for corresponding quantities to make their relationship obvious.
\begin{definition}[Set of Messages, $\mathcal{W}$] Define $\mathcal{W}=\{W_1, W_2, \cdots, W_K\}$ as the set of $K$ independent messages, each of size $L_w$ bits. 
\begin{eqnarray}
&& H(W_1, \cdots, W_K) = H(W_1) + \cdots + H(W_K),\\
&& L_w=H(W_1) = \cdots = H(W_K).
\end{eqnarray}
\end{definition}
\begin{definition}[Sets of Answers, $\mathcal{X}$, $\mathcal{X}^{[1:N]}$, Upload Cost] 
Define sets $\mathcal{X}^{[n]}=\{X_1^{[n]}, X_2^{[n]}, \cdots, X^{[n]}_{M_n}\}$ containing all possible answers from Database $n$, $n\in[1:N]$, such that all answers have the same size, $L_x$.
\begin{align*}
L_x&=H(X_m^{[n]}), &&\forall n\in[1:N], m\in[1:M_n].
\end{align*}
The upload cost for Database $n$, is defined to be $\log(M_n)$  for all $n\in[1:N]$. Furthermore, define
\begin{align}
\mathcal{X}&=\bigcup_{n\in[1:N]}\mathcal{X}^{[n]}
\end{align}
as the set of all answers.
\end{definition}
Note that we assume all answers have the same size. Under `maximum download' formulation of PIR, there is no loss of generality in this assumption because the rate of a PIR scheme is limited only by the largest possible download (answer) from any database for any query. If different possible answers have different lengths, then smaller answers can be padded with useless information to match the length of the biggest answer (maximum download).

\begin{definition}[IR ($\mathcal{A}, \mathcal{S}_{[1:K]}$)] \label{def:9} An $N$-query Information Retrieval scheme is comprised of a mapping $\mathcal{A}$ from the set of messages $\mathcal{W}$ to the sets of answers $\mathcal{X}^{[1:N]}$, and $K$  non-empty sets, $\mathcal{S}_{k}$, $k\in[1:K]$, called decoding supersets. Elements of the decoding supserset $\mathcal{S}_{k}$, are called decoding sets for the message $W_k$. Each decoding set for $W_k$ is of the form $S=\{X^{[1]}_{q_1}, X^{[2]}_{q_2}, \cdots, X^{[N]}_{q_N}\}$ with $q_n\in[1:M_n], \forall n\in[1:N]$ such that 
\begin{align}
S\in \mathcal{S}_k\Rightarrow H(W_k\mid S)&=0, &\forall k\in[1:K]. && \mbox{\normalfont[Correctness]}\label{eq:correct}
\end{align}
\end{definition}

The parameter $N$ is recognized as the number of databases. 
The  elements of the decoding set, $X_{q_n}^{[n]}$ represent what is requested by the user from the $n^{th}$ database, i.e., the query sent to Database $n$ is $q_n$ and the answer received from Database $n$ is $X_{q_n}^{[n]}$. If the desired message is $W_\theta$, then a decoding set is chosen  from $\mathcal{S}_\theta$.  Condition \eqref{eq:correct} is called the `correctness' condition, because it guarantees that the message can be decoded correctly from the answers received from all $N$ databases. Definition \ref{def:9} is useful only as a baseline  for introducing more interesting forms of information retrieval. The most interesting for our purpose is  perfectly private information retrieval, or simply PIR.
\begin{definition}[Perfectly Private Information Retrieval (PIR$_{\max}$)] A  PIR scheme is an  $N$-query Information Retrieval scheme with a distribution defined on the elements of each decoding superset (so we have $K$ distributions, one for each decoding superset), such that for all $n\in[1:N]$, and for all $k,k'\in[1:K]$ the conditional distribution of $q_n$ given $S\in \mathcal{S}_k$ is identical to the conditional distribution of $q_n$ given $S\in\mathcal{S}_{k'}$.
\begin{align}
\mbox{\normalfont Prob}(q_n=q\mid S\in\mathcal{S}_k)&=\mbox{\normalfont Prob}(q_n=q\mid S\in\mathcal{S}_{k'}), &\forall k,k'\in[1:K], n\in[1:N], \forall q\in[1:M_n].\label{eq:privacy}
\end{align}
\end{definition}

Equation \eqref{eq:privacy} ensures perfect privacy for the desired message index, because the query sent to any database has the same distribution regardless of the desired message index.
It is useful to also define a broader class of $N$-query Information Retrieval schemes, called Repudiative Information Retrieval (RIR), which includes PIR as a special case.
\begin{definition}[Repudiative Information Retrieval (RIR$_{\max}$)] \label{def:11} An RIR scheme is an $N$-query Information Retrieval scheme  such that every possible answer from every database appears in at least one of the decoding sets of every $\mathcal{S}_k, k\in[1:K].$
\begin{eqnarray}
\forall n\in[1:N], \forall m \in [1:M_n], ~\forall k \in [1:K], ~\exists S\in \mathcal{S}_{k} ~\mbox{such that}~X^{[n]}_{m} \in S.
\end{eqnarray}
\end{definition}
\begin{definition}[Rate and Capacity] The rate of an $N$-query information retrieval scheme is defined as
\begin{align}
R&=\frac{L_w}{NL_x} \label{eq:rpir}
\end{align}
and the supremum of $R$ values for a class of information retrieval schemes is called the capacity of that class.
\end{definition}

\subsection{Connection between ULDCs and RIR$_{\max}$}
It is well known that LDCs and PIR schemes are closely related \cite{YekhaninPhd}. Comparing preceding definitions for LDCs with locality $N$ and $N$-query information retrieval,  it is evident that source symbols correspond to messages, coded symbols correspond to answers, code length corresponds to total upload cost, SLDCs correspond to PIR$_{\max}$, the relaxation to ULDCs correspond to the relaxation to RIR$_{\max}$, and the decoding sets, rates and capacity expressions for both settings are similar as well. However, a closer look also reveals clear differences. For example, answers are partitioned into $\mathcal{X}^{[n]}, n\in[1:N]$, whereas no such partitioning is invoked for coded symbols. While both SLDCs and PIR$_{\max}$ impose  additional constraints on the decoding sets,  the two constraints are not  equivalent. These  distinctions often do not matter much in practice, indeed most PIR$_{\max}$ schemes produce SLDCs and most constructions of SLDCs are obtained from PIR$_{\max}$ schemes. Nevertheless, the  distinctions pose  difficulties in translating theoretical results between the two problems. For our purpose, the precise connection (obvious from the preceding definitions) that allows us to connect our results across the two settings is between ULDCs and RIR$_{\max}$, as stated below.
\begin{observation}\label{thm:LDCPIR}
The set of all answers $\mathcal{X}$ from an RIR$_{\max}$ scheme with message set $\mathcal{W}$, $N$ databases, upload costs $\log(M_{[1:N]})$, decoding supersets $\mathcal{S}_{[1:K]}$ and rate $R$, constitutes a ULDC with set of source symbols $\mathcal{W}$, coded symbols $\mathcal{X}$, locality $N$, code length $M=\sum_{n\in[1:N]}M_n$, decoding supersets $\mathcal{S}_{[1:K]}$, and symbol rate $R_s=NR$.
\end{observation}

Given the translation from RIR$_{\max}$ to ULDCs, one might be interested in the other direction, i.e., the translation from ULDCs to RIR$_{\max}$, which is also possible, although in general less efficient. For example, by choosing the sets of answers $\mathcal{X}^{[n]}, n\in[1:N]$, to be \emph{each} identical to the set of coded symbols $\mathcal{X}$ of a ULDC, an RIR$_{\max}$ scheme is trivially obtained. This is less efficient because of the \emph{expansion} by the factor $N$, i.e., the value of $\sum_{n\in[1:N]}M_n$ for the resulting RIR$_{\max}$ scheme is $N$ times  larger than the code length $M$ of the ULDC. Note that no such expansion occurs in the reverse direction. Interestingly, as illustrated in Figure \ref{fig:exp} through an example, an expansion by a factor of $N$ is \emph{necessary} in some cases when translating a ULDC into an RIR$_{\max}$ scheme.
\begin{figure}[!h]
\begin{center}
\noindent\begin{tikzpicture}[scale=0.81]
\node (X1) at (0,3) [draw,thick,minimum width=1cm,minimum height=0.75cm, fill=blue!7] { $X_1$};
\node (X2) at (3,3) [draw,thick,minimum width=1cm,minimum height=0.75cm, fill=blue!7] {$X_2$};
\node (X3) at (0,1) [draw,thick,minimum width=1cm,minimum height=0.75cm, fill=blue!7] {$X_3$};
\node (X4) at (3,1) [draw,thick,minimum width=1cm,minimum height=0.75cm, fill=blue!7] {$X_4$};
\draw[very thick, black!30!green] (X1)--(X4);
\draw[very thick, black!30!green] (X2)--(X3);
\draw[very thick, blue] (X1)--(X2);
\draw[very thick, blue] (X3)--(X4);
\draw[very thick, red] (X1)--(X3);
\draw[very thick, red] (X2)--(X4);
\node (def) at (11.5,2){$\begin{array}{l}
W_1 = (a_1,a_2,a_3,a_4), ~W_2 = (b_1, b_2,b_3, b_4), ~W_3 = (c_1, c_2, c_3, c_4)\\
X_1 = (a_1, a_2, b_1, b_2, c_1, c_2), ~~ X_2 = (a_3, a_4, b_1, b_3, c_1, c_3) \\
X_3 = (a_1, a_3, b_3, b_4, c_2, c_4), ~~ X_4 = (a_2, a_4, b_2, b_4, c_3, c_4)
\end{array}
$};
\end{tikzpicture}
\end{center}
\caption{\small \it A ULDC (also an SLDC) with locality $N=2$ that encodes $K=3$ source symbols with  $L_w=4$ bits each, $W_1, W_2, W_3$, into $M=6$ coded symbols, $X_1, X_2, X_3, X_4$,  with $L_x=6$ bits each. The decoding sets for $W_1, W_2, W_3$ are comprised of pairs of coded symbols connected by blue, red, and green edges, respectively. It is easy to see that the only RIR$_{\max}$ scheme that can be constructed from this ULDC is with answer sets $\{X_1, X_2, X_3, X_4\}$ replicated at the $N=2$ databases. Therefore, the total number of answers is $8$, $N=2$ times the ULDC length, i.e., we have an expansion by a factor of $N=2$. }\label{fig:exp}
\end{figure}

Note that since ULDCs and RIR$_{\max}$ are relaxations of SLDCs and PIR$_{\max}$, respectively, impossibility results (converse arguments) for ULDCs and RIR$_{\max}$ apply to SLDCs and PIR$_{\max}$ automatically, while achievable schemes for the SLDCs and PIR$_{\max}$ apply  automatically to ULDCs and RIR$_{\max}$. These inclusions will be useful to prove our main results, presented in the next section.

\section{Main Results}\label{sec:main}
\subsection{Capacity Results}
Our first set of results are capacity characterizations. 
Given $K$ source symbols, code length $M$, and locality $N$, let $C_{\mbox{\tiny SLDC}}(N, K, M)$ and $C_{\mbox{\tiny ULDC}}(N,K,M)$ denote the capacity for the class of SLDCs and ULDCs respectively. Our first result characterizes the maximum possible capacity of a ULDC given the  the locality $N$ and the number of source symbols $K.$ The maximum is over all possible codeword lengths $M$.
\begin{theorem}\label{thm:capacity}
\begin{align}
 C_{\mbox{\tiny ULDC}}^*(N,K)&\define \max_{M\in\mathbb{N}} C_{\mbox{\tiny ULDC}}(N,K,M)\notag\\
  &= N\left(1 +1/N + 1/{N^2} + \cdots +1/{N^{K-1}}\right)^{-1}.
\end{align}
\end{theorem}
The expression for $C_{\mbox{\tiny ULDC}}^*(N,K)$ is reminiscent of the capacity of PIR \cite{Sun_Jafar_PIR}. Indeed, since the capacity achieving PIR schemes in \cite{Sun_Jafar_PIR} naturally  produce  SLDCs, and all SLDCs are also ULDCs, the achievability argument is directly implied. However, since ULDCs are a more general class of objects than the LDCs  produced by PIR schemes, the converse from \cite{Sun_Jafar_PIR} does not apply.  Instead, a new combinatorial converse proof is presented for Theorem \ref{thm:capacity} in Section \ref{sec:capacityconverse}. As an immediate corollary, we settle the corresponding question for SLDCs as well. 

\begin{corollary}\label{cor:capacitySLDC}
\begin{align}
 C_{\mbox{\tiny SLDC}}^*(N,K)&\define \max_{M\in\mathbb{N}} C_{\mbox{\tiny SLDC}}(N,K,M)\notag\\
  &= N\left(1 +1/N + 1/{N^2} + \cdots +1/{N^{K-1}}\right)^{-1}.
\end{align}
\end{corollary}
The achievability argument for Corollary \ref{cor:capacitySLDC} follows from the capacity achieving PIR schemes in \cite{Sun_Jafar_PIR} (note that Corollary \ref{cor:length}, to be presented in the next subsection, also contains a capacity achieving SLDC). The converse follows from Theorem \ref{thm:capacity} as SLDCs are special cases of ULDCs.

As another corollary, the capacity of RIR$_{\max}$ is shown to be the same as the capacity of PIR$_{\max}$. 
\begin{corollary}\label{cor:capacityRIR}
\begin{align}
 C_{\mbox{\tiny RIR$_{\max}$}}(N,K)&= \left(1 +1/N + 1/{N^2} + \cdots +1/{N^{K-1}}\right)^{-1}= C_{\mbox{\tiny PIR$_{\max}$}}(N,K)= C_{\mbox{\tiny PIR$_{\small ave}$}}(N,K).
\end{align}
\end{corollary}
The achievability for Corollary \ref{cor:capacityRIR} follows because PIR$_{\max}$ schemes are special cases of RIR$_{\max}$ schemes and capacity achieving PIR$_{\max}$ schemes are available from \cite{Sun_Jafar_PIR}. The converse follows from Observation \ref{thm:LDCPIR} and Theorem \ref{thm:capacity}. That is, the rate of any RIR$_{\max}$ scheme must be no higher than  $C_{\mbox{\tiny RIR$_{\max}$}}(N,K)$, otherwise by Observation \ref{thm:LDCPIR} we will have a ULDC that has a rate higher than $C_{\mbox{\tiny ULDC}}^*(N,K)$, contradicting Theorem \ref{thm:capacity}.

\subsection{Optimal Code Length and Upload Cost Results}
The next set of results concerns minimum code lengths and minimum upload costs. We first show that given $N, K$, the minimum code length $M$ of  ULDCs for which the capacity takes its maximum value (maximum over all $M$), is $N^K$.
\begin{theorem}\label{thm:length}
\begin{align}
\min\{M\mid C_{\mbox{\tiny ULDC}}(N,K,M) = C_{\mbox{\tiny ULDC}}^*(N,K)\}&=N^{K}.
\end{align}
\end{theorem}

For the converse, we prove that any capacity achieving ULDCs must have length $M \geq N^K$. The proof is presented in Section \ref{sec:length}. Since SLDCs are special cases of ULDCs, the converse also applies to SLDCs. 
For the achievability, we provide a construction of a capacity achieving SLDC with length $M = N^K$. The proof is presented in Section \ref{sec:length_ach}. Since every SLDC is also a ULDC, the achievability applies also to ULDCs. Thus, we immediately have the following corollary for SLDCs.
\begin{corollary}\label{cor:length}
\begin{align}
\min\{M\mid C_{\mbox{\tiny SLDC}}(N,K,M) = C_{\mbox{\tiny SLDC}}^*(N,K)\}&=N^{K}.
\end{align}
\end{corollary}

\begin{corollary}\label{cor:rir}
The minimum upload cost of a capacity achieving RIR$_{\max}$ scheme with $K$ messages and $N$ databases is $(K-1)\log(N)$ per database.
\end{corollary}

\begin{corollary}\label{cor:pir}
The minimum upload cost of a capacity achieving PIR$_{\max}$ scheme with $K$ messages and $N$ databases is $(K-1)\log(N)$ per database.
\end{corollary}

The proofs of Corollaries \ref{cor:rir} and \ref{cor:pir} are presented in Section \ref{sec:pir}.

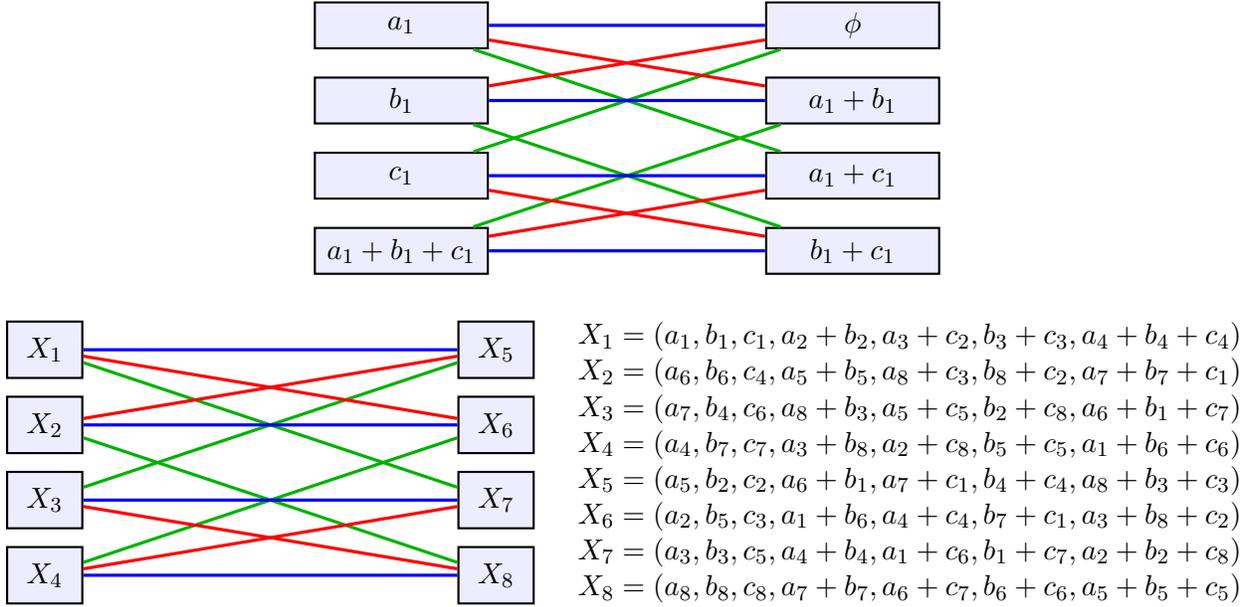
\begin{figure}[h]
\begin{center}
\begin{tikzpicture}[baseline]
\node (X1) at (0,3) [draw,thick, minimum width=2.3cm,minimum height=0.61cm, fill=blue!7] { $a_1$};
\node (X2) at (0,2) [draw,thick, minimum width=2.3cm,minimum height=0.61cm, fill=blue!7] {$b_1$};
\node (X3) at (0,1) [draw,thick,minimum width=2.3cm,minimum height=0.61cm, fill=blue!7] {$c_1$};
\node (X4) at (0,0) [draw,thick,minimum width=2.3cm,minimum height=0.61cm, fill=blue!7] {$a_1+b_1+c_1$};
\node (X5) at (6,3) [draw,thick,minimum width=2.3cm,minimum height=0.61cm, fill=blue!7] {$\phi$};
\node (X6) at (6,2) [draw,thick, minimum width=2.3cm,minimum height=0.61cm, fill=blue!7] {$a_1+b_1$};
\node (X7) at (6,1) [draw,thick,minimum width=2.3cm,minimum height=0.61cm, fill=blue!7] { $a_1+c_1$};
\node (X8) at (6,0) [draw,thick,minimum width=2.3cm,minimum height=0.61cm, fill=blue!7] {$b_1+c_1$};
\draw[very thick, black!30!green] (X1)--(X7);
\draw[very thick, black!30!green] (X3)--(X5);
\draw[very thick, black!30!green] (X2)--(X8);
\draw[very thick, black!30!green] (X4)--(X6);
\draw[very thick, blue] (X1)--(X5);
\draw[very thick, blue] (X2)--(X6);
\draw[very thick, blue] (X3)--(X7);
\draw[very thick, blue] (X4)--(X8);
\draw[very thick, red] (X1)--(X6);
\draw[very thick, red] (X2)--(X5);
\draw[very thick, red] (X3)--(X8);
\draw[very thick, red] (X4)--(X7);
\end{tikzpicture}
\end{center}

\noindent\begin{tikzpicture}[baseline]
\node (X1) at (0,3) [draw,thick,minimum width=1cm,minimum height=0.75cm, fill=blue!7] { $X_1$};
\node (X2) at (0,2) [draw,thick,minimum width=1cm,minimum height=0.75cm, fill=blue!7] {$X_2$};
\node (X3) at (0,1) [draw,thick,minimum width=1cm,minimum height=0.75cm, fill=blue!7] {$X_3$};
\node (X4) at (0,0) [draw,thick,minimum width=1cm,minimum height=0.75cm, fill=blue!7] {$X_4$};
\node (X5) at (6,3) [draw,thick,minimum width=1cm,minimum height=0.75cm, fill=blue!7] {$X_5$};
\node (X6) at (6,2) [draw,thick,minimum width=1cm,minimum height=0.75cm, fill=blue!7] {$X_6$};
\node (X7) at (6,1) [draw,thick,minimum width=1cm,minimum height=0.75cm, fill=blue!7] { $X_7$};
\node (X8) at (6,0) [draw,thick,minimum width=1cm,minimum height=0.75cm, fill=blue!7] {$X_8$};
\draw[very thick, black!30!green] (X1)--(X7);
\draw[very thick, black!30!green] (X3)--(X5);
\draw[very thick, black!30!green] (X2)--(X8);
\draw[very thick, black!30!green] (X4)--(X6);
\draw[very thick, blue] (X1)--(X5);
\draw[very thick, blue] (X2)--(X6);
\draw[very thick, blue] (X3)--(X7);
\draw[very thick, blue] (X4)--(X8);
\draw[very thick, red] (X1)--(X6);
\draw[very thick, red] (X2)--(X5);
\draw[very thick, red] (X3)--(X8);
\draw[very thick, red] (X4)--(X7);
\node (def) at (11.5,1.5){$\begin{array}{l}
X_1=(a_1,b_1,c_1,a_2+b_2,a_3+c_2,b_3+c_3, a_4+b_4+c_4)\\
X_2=(a_6,b_6,c_4,a_5+b_5,a_8+c_3,b_8+c_2, a_7+b_7+c_1)\\
X_3=(a_7,b_4,c_6,a_8+b_3,a_5+c_5,b_2+c_8, a_6+b_1+c_7)\\
X_4=(a_4,b_7,c_7,a_3+b_8,a_2+c_8,b_5+c_5, a_1+b_6+c_6)\\
X_5=(a_5,b_2,c_2,a_6+b_1,a_7+c_1,b_4+c_4, a_8+b_3+c_3)\\
X_6=(a_2,b_5,c_3,a_1+b_6,a_4+c_4,b_7+c_1, a_3+b_8+c_2)\\
X_7=(a_3,b_3,c_5,a_4+b_4,a_1+c_6,b_1+c_7, a_2+b_2+c_8)\\
X_8=(a_8,b_8,c_8,a_7+b_7,a_6+c_7,b_6+c_6, a_5+b_5+c_5)
\end{array}
$};
\end{tikzpicture}
\caption{\it \small Shown at the top is a capacity achieving, upload optimal PIR$_{\mbox{\footnotesize ave}}$ scheme for $K=3$ messages, $N=2$ databases from \cite{Tian_Sun_Chen_Upload}. At the bottom is the corresponding capacity achieving, upload optimal PIR$_{\max}$ scheme from this work. The messages are denoted by $W_1=a_{[1:L_w]}, W_2=b_{[1:L_w]}, W_3=c_{[1:L_w]}$, in both cases, with $L_w=1$ for PIR$_{\mbox{\footnotesize ave}}$ and $L_w=8$ for PIR$_{\max}$. Nodes in the left column are all possible answers from Database $1$, and the nodes in the right column are all possible answers from Database $2$. In both cases, $W_1$ can be retrieved from pairs of nodes connected by blue edges, $W_2$ from red edges and $W_3$ from green edges.}\label{fig:maxaveup}
\end{figure}

It is already known from \cite{Sun_Jafar_PIR} that the capacity of PIR$_{\max}$ is the same as the capacity of PIR$_{\mbox{\footnotesize ave}}$. Surprisingly, based on Corollary \ref{cor:pir} and the results in \cite{Tian_Sun_Chen_Upload}, it turns out that the minimum upload cost for PIR$_{\max}$ is also the same as the minimum upload cost of PIR$_{\mbox{\small ave}}$. Note that any capacity achieving, upload optimal PIR$_{\max}$ scheme is also a capacity achieving, upload optimal PIR$_{\mbox{\footnotesize ave}}$ scheme. However, the reverse direction is not true.  This is evident from Figure \ref{fig:maxaveup} which shows capacity achieving and upload optimal schemes for both settings.

The PIR$_{\mbox{\footnotesize ave}}$ scheme shown in Figure \ref{fig:maxaveup} uses message size $L_w=1$ bit and achieves an average download of $L_w$ from Database $1$, and $\frac{3}{4}L_w=3/4$ from Database $2$, for total average download of $\frac{7}{4}L_w$, so its rate is $4/7$, the capacity for this setting. Note that this is because with probability $1/4$ nothing is downloaded from Database $2$. However, the maximum download for this scheme is $L_w$ per database which is not optimal. Therefore, using the answers from this scheme directly to produce an LDC would result in an LDC with $L_x=L_w$, which is  not capacity achieving. On the other hand, the PIR$_{\max}$ scheme shown in Figure \ref{fig:maxaveup} uses message size $L_w=8$ bits, and achieves constant, maximum, and average download of $\frac{7}{8}L_w=7$ bits from each database, for a total download of $\frac{7}{4}L_w$, so its rate is also $4/7$, same as the capacity for this setting. This is a stronger capacity achieving scheme because not only is it capacity achieving and upload optimal for PIR$_{\max}$ but also it is capacity achieving and upload optimal for PIR$_{\mbox{\footnotesize ave}}$. Furthermore, the same scheme gives us a minimum length capacity achieving ULDC, a minimum length capacity achieving SLDC, as well as a capacity achieving and upload optimal scheme for RIR$_{\max}$. Note that the upload optimal PIR$_{\max}$ scheme cannot be obtained simply from a time-sharing argument that symmetrizes the upload optimal PIR$_{\mbox{\footnotesize ave}}$ scheme, because the time-sharing argument increases the upload cost. Instead, this powerful scheme, which gets even more sophisticated for larger number of messages and databases,  is obtained by a special construction specified in Section \ref{sec:length_ach}.

\section{Converse Proof of Theorem \ref{thm:capacity}}\label{sec:capacityconverse}
Let us start with a simple yet extremely useful lemma.
\begin{lemma}\label{lemma:dec}
Let  $S \in \mathcal{S}_k$ be an arbitrary decoding set of $W_k$.
Consider an arbitrary subset of $[1:K]$, denoted by $\mathcal{J}$, such that $k \notin \mathcal{J}$. Then for any element $X_s$ in $S$, we have
\begin{align}
\sum_{X_i \in S} H(X_i | W_\mathcal{J})& \geq L_w + H(X_{s} | W_{\{k\} \cup \mathcal{J}}), &\forall X_s \in S. \label{eq:lemmadec}
\end{align}
\end{lemma}
{\it Proof:} 
\begin{eqnarray}
\sum_{X_i \in S} H(X_i | W_\mathcal{J}) &\geq& H(S | W_\mathcal{J}) \label{eq:l2}\\
&\overset{(a)}{=}& H(S, W_k | W_\mathcal{J}) \\
&\overset{(\ref{h1})}{=}& H(W_k) + H(S | W_k,W_\mathcal{J}) \\
&\overset{(\ref{h2})}{\geq}& L_w + H(X_{s}| W_{\{k\}\cup \mathcal{J}}) \label{eq:l1}
\end{eqnarray}
where $(a)$ follows from the fact that $S$ is a decoding set of $W_k$, so from $S$, we may decode $W_k$. The last step is due to the assumption that $X_s \in S$. $\hfill\QED$

{\it Remark: Lemma \ref{lemma:dec} states that the amount of information contained in any decoding set of a source symbol is no less than the entropy of that source symbol plus the entropy of any coded symbol from the decoding set conditioned on that source symbol (i.e., interference about other source symbols).}

\bigskip
The rest of the proof follows from invoking Lemma \ref{lemma:dec} for a carefully chosen sequence of decoding sets and a permutation of the $K$ source symbols. Consider an arbitrary permutation of $[1:K]$, $\pi$ such that $(1,2,\cdots, K)$ is mapped to $(\pi_1, \pi_2, \cdots, \pi_K)$. 

\begin{figure}[h]
\begin{center}
\includegraphics[width= 6.0in]{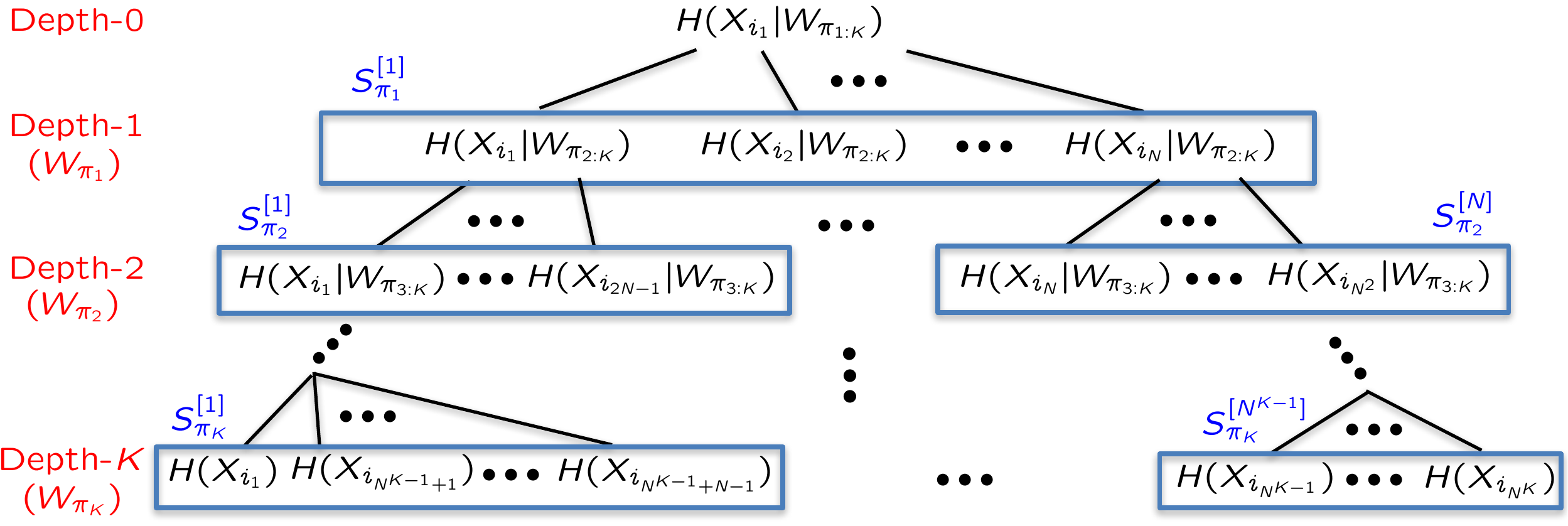}
\caption{\small The full $N$-ary tree with depth $K$ containing all coded symbols and decoding sets that appear in the converse proof. The indices of coded symbols are labelled lexicographically from the root to the leaf nodes (they are not necessarily distinct).
}
\label{fig:decodetree}
\end{center}
\end{figure}

The decoding sets and coded symbols involved in the converse proof are constructed following a full $N$-ary tree with depth $K$ (see Figure \ref{fig:decodetree}). At depth-$k, k \in [1:K]$, there are $N^{k-1}$ decoding sets (not necessarily distinct) of the source symbol $W_{\pi_k}$. Specifically, we start from the root, where we pick an arbitrary coded symbol, $X_{i_1}$. Because the LDC is universal, $X_{i_1}$ can be used to decode $W_{\pi_1}$, with another $N-1$ symbols (denoted as $X_{i_2}, \cdots, X_{i_N}$). These $N$ symbols form the depth-1 nodes and this decoding set is denoted as ${S}_{\pi_1}^{[1]}$. The remaining procedure is similar, where for each node at depth-$(k-1)$, we find a decoding set of the source symbol $W_{\pi_k}$ that contains it and these decoding sets appear at depth-$k$. Finally, at depth-$K$, we have $N^{K-1}$ decoding sets of the source symbol $W_{\pi_K}$. When referring to a node in the full $N$-ary tree, we may use either the content (i.e., the entropy term) or the $X_{i}$ value (called the node \emph{label}).

\begin{example}
To illustrate the construction of the full $N$-ary tree, we consider an example of a ULDC as shown in Figure \ref{fig:uldctree}. For one possible construction of the full binary tree, we set the permutation $\pi$ as the identity permutation and pick $X_1$ as the root node. To find the depth-1 nodes, we pick any decoding set of $W_1$ that contains $X_1$, say $\{X_1, X_2\} \triangleq S_1^{[1]}$, so that the depth-1 nodes are $H(X_1|W_2, W_3)$ and $H(X_2|W_2, W_3)$. Next, we find the depth-2 nodes. Consider the two depth-1 nodes and for each of them, we pick any decoding set of $W_2$ that contains the coded symbol in the depth-1 node. For the first depth-1 node $H(X_1|W_2, W_3)$, we only have 1 decoding set that contains $X_1$ (note that there must exist one as the LDC is universal), so $S_2^{[1]} = \{X_1, X_2\}$. For the second depth-1 node $H(X_2|W_2, W_3)$, we have 2 decoding sets that contain $X_2$ and we may choose either one,  say we choose $\{X_2, X_3\} \triangleq S_2^{[2]}$. We have now found the 4 depth-2 nodes, as $H(X_1|W_3), H(X_2|W_3), H(X_2|W_3)$, and $H(X_3|W_3)$, where the first two nodes are from $S_2^{[1]}$ and the last two nodes are from $S_2^{[2]}$. Note that the nodes at the same depth are not necessarily distinct, e.g., $X_2$ appears twice\footnote{However, for any ULDC to achieve the capacity, the nodes from the same depth must be distinct. We refer to the proof of Theorem \ref{thm:length} for the justification of this distinctness property. Therefore, it follows that this ULDC does not achieve the capacity, verified by noting that the symbol rate is $R = L_w/L_x = 1$ while the capacity is $C_{\mbox{\tiny ULDC}}^*(N=2,K=2) = 4/3$.} at depth-2. Finally, we consider the depth-$K$ (depth-3) nodes. For each one of the depth-2 nodes, we find a decoding set of $W_3$ that contains it, e.g., $S_3^{[1]} = \{X_1, X_3\}, S_3^{[2]} = \{X_2, X_3\}, S_3^{[3]} = \{X_2, X_4\}, S_3^{[4]} = \{X_3, X_2\}$, then the depth-3 nodes are $H(X_1), H(X_3), H(X_2), H(X_3), H(X_2), H(X_4), H(X_3), H(X_2)$, where sequentially every 2 nodes form a decoding set of $W_3$. The construction of the full binary tree is now complete.
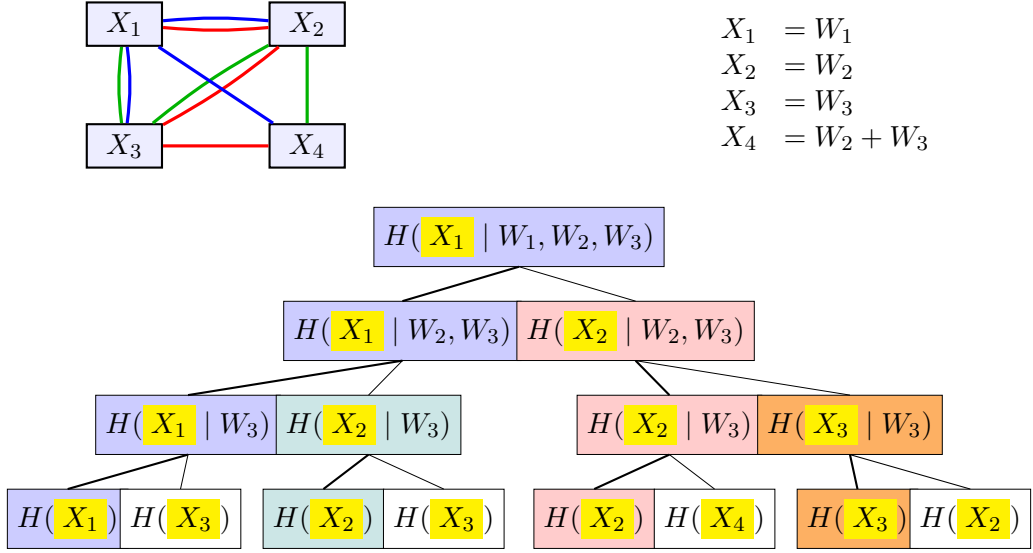
\begin{figure}
\begin{center}
\begin{tikzpicture}[scale=0.81]
\node (X1) at (0,3) [draw,thick,minimum width=1cm,minimum height=0.5cm, fill=blue!7] { $X_1$};
\node (X2) at (3,3) [draw,thick,minimum width=1cm,minimum height=0.5cm, fill=blue!7] {$X_2$};
\node (X3) at (0,1) [draw,thick,minimum width=1cm,minimum height=0.5cm, fill=blue!7] {$X_3$};
\node (X4) at (3,1) [draw,thick,minimum width=1cm,minimum height=0.5cm, fill=blue!7] {$X_4$};

\draw[very thick, bend left=5, blue] (X1)edge(X2);
\draw[very thick, bend right=5, red] (X1)edge(X2);
\draw[very thick, bend left=7, blue] (X1)edge(X3);
\draw[very thick, bend right=7, black!30!green] (X1)edge(X3);
\draw[very thick, bend left=5, red] (X2)edge(X3);
\draw[very thick, bend right=5, black!30!green] (X2)edge(X3);
\draw[very thick, blue] (X1)--(X4);
\draw[very thick, black!30!green] (X2)--(X4);
\draw[very thick, red] (X3)--(X4);
\node (def) at (11.5,2){$\begin{array}{ll}
X_1&=W_1\\
X_2&=W_2\\
X_3&=W_3\\
X_4&=W_2+W_3
\end{array}
$};
\end{tikzpicture}\\
$~$\\
\begin{tikzpicture}
\node (A1) at (0,3.75) [draw, ultra thin, fill=blue!20, text=black] { $H({\colorbox{yellow}{$X_1$}}\mid W_1, W_2, W_3)$};

\node (B1) at (-1.55,2.5) [draw,  ultra thin, fill=blue!20, text=black] { $H({\colorbox{yellow}{$X_1$}}\mid W_2, W_3)$};
\node (B2) at (1.55,2.5)  [draw, ultra thin, fill=red!20, text=black]{ $H({\colorbox{yellow}{$X_2$}}\mid W_2, W_3)$};

\node (C1) at (-4.4,1.25)[draw,  ultra thin, fill=blue!20, text=black]  { $H({\colorbox{yellow}{$X_1$}}\mid W_3)$};
\node (C2) at (-2,1.25) [draw,  ultra thin, fill=teal!20, text=black] { $H({\colorbox{yellow}{$X_2$}}\mid W_3)$};

\node (C3) at (2,1.25) [draw,  ultra thin, fill=red!20, text=black] { $H({\colorbox{yellow}{$X_2$}}\mid W_3)$};
\node (C4) at (4.4,1.25) [draw,  ultra thin, fill=black!1!orange!61, text=black] { $H({\colorbox{yellow}{$X_3$}}\mid W_3)$};

\node (D1) at (-6,0)[draw,  ultra thin, fill=blue!20, text=black]  { $H({\colorbox{yellow}{$X_1$}})$};
\node (D2) at (-4.5,0)[draw,  ultra thin, fill=white, text=black]  { $H({\colorbox{yellow}{$X_3$}})$};

\node (D3) at (-2.6,0) [draw,  ultra thin, fill=teal!20, text=black] { $H({\colorbox{yellow}{$X_2$}})$};
\node (D4) at (-1,0) [draw,  ultra thin, fill=white, text=black] { $H({\colorbox{yellow}{$X_3$}})$};

\node (D5) at (1,0) [draw,  ultra thin, fill=red!20, text=black] { $H({\colorbox{yellow}{$X_2$}})$};
\node (D6) at (2.6,0)[draw,  ultra thin, fill=white, text=black]  { $H({\colorbox{yellow}{$X_4$}})$};

\node (D7) at (4.5,0) [draw,  ultra thin, fill=black!1!orange!61, text=black] { $H({\colorbox{yellow}{$X_3$}})$};
\node (D8) at (6,0)[draw,  ultra thin, fill=white, text=black]  { $H({\colorbox{yellow}{$X_2$}})$};

\draw[thick, black] (A1.south)--(B1.north);
\draw[ultra thin, black] (A1.south)--(B2.north);
\draw[thick, black] (B1.south)--(C1.north);
\draw[ultra thin, black] (B1.south)--(C2.north);
\draw[thick, black] (B2.south)--(C3.north);
\draw[ultra thin, black] (B2.south)--(C4.north);
\draw[thick, black] (C1.south)--(D1.north);
\draw[ultra thin, black] (C1.south)--(D2.north);
\draw[thick, black] (C2.south)--(D3.north);
\draw[ultra thin, black] (C2.south)--(D4.north);
\draw[thick, black] (C3.south)--(D5.north);
\draw[ultra thin, black] (C3.south)--(D6.north);
\draw[thick, black] (C4.south)--(D7.north);
\draw[thin, black] (C4.south)--(D8.north);
\end{tikzpicture}
\end{center}
\caption{\it\small Shown at the top of the figure is a ULDC with locality $N=2$ that codes $K=3$ binary source symbols, $W_1, W_2, W_3$, into $M=4$ binary coded symbols, $X_1, X_2, X_3, X_4$. The decoding sets for $W_1, W_2, W_3$ are shown as pairs of coded symbols connected by blue, red, and green edges, respectively. At the bottom of the figure is one possible $N$-ary tree for this ULDC. Node labels are the $X_i$ values highlighted in yellow.}\label{fig:uldctree}
\end{figure}

\end{example}

{\it Remark: From this example, it is clear that there are many different ways to generate the full $N$-ary tree (e.g., the permutation can be chosen arbitrarily, the root node can be chosen arbitrarily, and when there are multiple qualified decoding sets, any one may be chosen). Interestingly, the converse proof works for any realization of the full $N$-ary tree.
}

\bigskip
For the converse proof, we start from the $N^{K-1}$ decoding sets of the source symbol $W_{\pi_K}$ at depth-$K$ and repeatedly apply Lemma \ref{lemma:dec} as we ascend the tree, and stop when we reach the root.
\begin{eqnarray}
N^K L_x&=& \sum_{n=1}^{N^{K-1}} \sum_{X_i \in {S}_{\pi_K}^{[n]}} H(X_i) \\
&\overset{(\ref{eq:lemmadec})}{\geq}& N^{K-1} L_w + \sum_{n=1}^{N^{K-2}}  \sum_{X_i \in {S}_{\pi_{K-1}}^{[n]}} H(X_{i} | W_{\pi_K}) \label{eq:p1}\\
&\overset{(\ref{eq:lemmadec})}{\geq}& N^{K-1} L_w + N^{K-2} L_w  + \sum_{n=1}^{N^{K-3}}  \sum_{X_i \in {S}_{\pi_{K-2}}^{[n]}} H(X_{i} | W_{\pi_{K-1:K}}) \\
&\overset{}{\geq}& \cdots \\
&\overset{(\ref{eq:lemmadec})}{\geq}& N^{K-1} L_w + \cdots + NL_w + \sum_{X_i\in {S}_{\pi_1}^{[1]}} H(X_{i} | W_{\pi_{2:K}})\\
&\overset{(\ref{eq:lemmadec})}{\geq}& N^{K-1} L_w + \cdots + NL_w + L_w + H(X_{i_1} | W_{\pi_{1:K}}) \label{eq:p2}\\
&\overset{}{\geq}& (N^{K-1} + \cdots + N + 1)L_w \label{eq:pf}
\end{eqnarray}
We obtain the final rate bound by rearranging terms.
\begin{eqnarray}
R_s = \frac{L_w}{L_x} &\leq& N\left(1+\frac{1}{N} + \cdots + \frac{1}{N^{K-1}}\right)^{-1}.
\end{eqnarray} $\hfill\QED$

\section{Proof of Theorem \ref{thm:length}: Converse}\label{sec:length}
We show that a capacity achieving ULDC has length at least $N^K$. 
For a set $\mathcal{K} \subset [1:K]$, denote its complement set as $\overline{\mathcal{K}}$ (i.e., the set of elements that are not in $\mathcal{K}$). We start by defining when two coded symbols contain the same information about a source symbol set.
\begin{definition}[Same information]
We say that two coded symbols $X_{i_1}, X_{i_2}$ contain the same information about a set of source symbols $W_{\mathcal{K}}$ if $H(X_{i_1} | X_{i_2}, W_{\overline{\mathcal{K}}}) = H(X_{i_2} | X_{i_1}, W_{\overline{\mathcal{K}}}) = 0$ and denote it as $X_{i_1} \overset{W_{\mathcal{K}}}{\simeq} X_{i_2}$.
\end{definition}

By definition, the same information operation is symmetric, i.e., if $X_{i_1} \overset{W_{\mathcal{K}}}{\simeq} X_{i_2}$, then $X_{i_2} \overset{W_{\mathcal{K}}}{\simeq} X_{i_1}$. Interestingly, the same information operation is also transitive. This is proved in the following lemma.
\begin{lemma} [Transitivity of same information] \label{lemma:trans}
If $X_{i_1} \overset{W_{\mathcal{K}}}{\simeq} X_{i_2}$ and $X_{i_2} \overset{W_{\mathcal{K}}}{\simeq} X_{i_3}$, then $X_{i_1} \overset{W_{\mathcal{K}}}{\simeq} X_{i_3}$.
\end{lemma}
{\it Proof:} We show that $H(X_{i_1} | X_{i_3}, W_{\overline{\mathcal{K}}}) = 0$, and the proof of $H(X_{i_3} | X_{i_1}, W_{\overline{\mathcal{K}}}) = 0$ follows by symmetry.
\begin{eqnarray}
H(X_{i_1} | X_{i_3}, W_{\overline{\mathcal{K}}}) &=& H(X_{i_1} | X_{i_2}, X_{i_3}, W_{\overline{\mathcal{K}}}) + I(X_{i_1}; X_{i_2} | X_{i_3}, W_{\overline{\mathcal{K}}}) \\
&=& H(X_{i_1} | X_{i_2}, X_{i_3}, W_{\overline{\mathcal{K}}}) + H(X_{i_2} | X_{i_3}, W_{\overline{\mathcal{K}}}) - H(X_{i_2} | X_{i_1}, X_{i_3}, W_{\overline{\mathcal{K}}}) \label{eq:dec0}\\
&=& 0
\end{eqnarray}
where in (\ref{eq:dec0}), the first term is zero because $X_{i_1} \overset{W_{\mathcal{K}}}{\simeq} X_{i_2}$ (i.e., $H(X_{i_1} | X_{i_2}, W_{\overline{\mathcal{K}}}) = 0$) and adding conditioning can not increase entropy and the last two terms are zero because $X_{i_2} \overset{W_{\mathcal{K}}}{\simeq} X_{i_3}$. $\hfill\QED$

Similarly, we define when two coded symbols contain distinct information about a single source symbol.
\begin{definition}[Distinct information]
We say that two coded symbols $X_{i_1}, X_{i_2}$ contain distinct information about the source symbol $W_k, k \in [1:K]$ if $H(X_{i_1} | X_{i_2}, W_{\overline{k}}) = H(X_{i_1} | W_{\overline{k}})$ and denote it as $X_{i_1} \overset{W_{k}}{\perp} X_{i_2}$.
\end{definition}

Next we distill properties of capacity achieving ULDCs.

\begin{lemma}[Properties of capacity achieving ULDC] \label{lemma:cap}
For capacity achieving ULDCs, we have
\begin{enumerate}
\item (Non-zero entropy property) $\forall i \in [1:M]$, $\forall k \in [1:K]$, $H(X_i | W_{\overline{k}}) \neq 0$.
\item For an arbitrary decoding set of $W_k, k \in [1:K]$, $S \in \mathcal{S}_k$,
\begin{enumerate}
\item (Same interference property) $\forall i_{1}, i_{2} \in {S}, \forall k' \neq k$, $X_{i_{1}} \overset{W_{k'}}{\simeq} X_{i_2}$. 
\item (Distinct desired information property) $\forall i_{1}, i_{2} \in {S}$, $X_{i_{1}} \overset{W_{k}}{\perp} X_{i_2}$. 
\item (Independence of coded symbols) $\forall i_{1}, i_{2} \in {S}$, $H(X_{i_1} | X_{i_2}) = H(X_{i_1})$.
\end{enumerate}
\item (Incompatibility of same and distinct information) There do not exist coded symbols $X_{i_1}, X_{i_2}$ and source symbol $W_k$ such that $X_{i_{1}} \overset{W_{k}}{\simeq} X_{i_2}$ and $X_{i_{1}} \overset{W_{k}}{\perp} X_{i_2}$.
\end{enumerate}
\end{lemma}

The proof of Lemma \ref{lemma:cap} is deferred to Section \ref{sec:lemmacap}.

{\it Remark: The idea of using properties on same interference and distinct information has appeared previously in \cite{Tian_Sun_Chen_Upload}, albeit within a restricted class of decomposable (e.g., linear) schemes. Here we develop them in the information theoretic sense (that works for any non-linear schemes). Further we treat same and distinct information as general mathematical operators and establish the transitivity of same information and incompatibility of same and distinct information.}

\bigskip
Equipped with the definitions and lemmas presented above, we are now ready for the proof, i.e., any capacity achieving ULDC must have length $M \geq N^K$. The proof idea is to consider a full $N$-ary tree (refer to Figure \ref{fig:decodetree}) that contains $N^K$ coded symbols and show that these coded symbols must be all distinct (so the length $M \geq N^K$). To this end, we show that if any two coded symbols are the same, then the ULDC can not achieve the capacity (as some properties established in Lemma \ref{lemma:cap} are violated).
To illustrate the idea in a simpler setting, let us start from an example with $N=2, K=3$.

\subsection{Example: $N=2, K=3$}
\begin{figure}[h]
\begin{center}
\includegraphics[width= 5.0in]{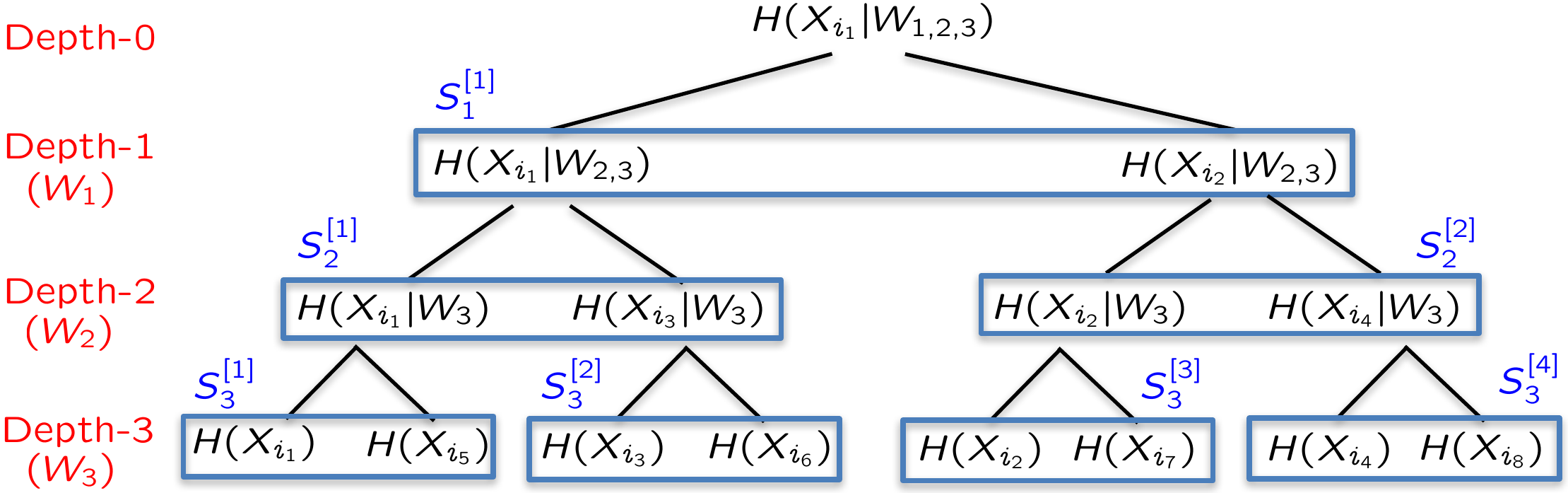}
\caption{\small The full binary tree with locality $N=2$ and $K=3$ messages.}
\label{fig:n2k3}
\end{center}
\end{figure}

We redraw the full binary tree with depth 3 in Figure \ref{fig:n2k3}, when the permutation is the identity permutation. There are $N^K = 8$ coded symbols (leaf nodes) involved, i.e., $X_{i_1}, \cdots, X_{i_8}$, and we show that they are all distinct, i.e., $X_{i_{j}}\neq X_{i_{l}}, \forall j, l \in [1:8], j \neq l$. This is proved by contradiction, i.e., if $X_{i_{j}} = X_{i_{l}}$, then the ULDC violates some property that must be satisfied by capacity achieving ULDCs.

We have 3 cases for the 2 leaf nodes $X_{i_{j}}, X_{i_{l}}$.
\begin{enumerate}
\item $X_{i_{j}}, X_{i_{l}}$ are siblings (i.e., $X_{i_{j}}, X_{i_{l}}$ have the same parent). For example, $X_{i_1}$ and $X_{i_5}$ are siblings. Now if $X_{i_1} = X_{i_5}$, we have $H(X_{i_1}|X_{i_5}) = 0$. Noting that $X_{i_1}, X_{i_5}$ form a decoding set of $W_3$, we apply the independence property of coded symbols (Property 2.(c)), and obtain $H(X_{i_1}) = H(X_{i_1}|X_{i_5}) = 0$, which contradicts  the fact that $H(X_{i_1}) = L_x \neq 0$ (as the code is capacity achieving). Therefore $X_{i_1}, X_{i_5}$ must be distinct.
\item $X_{i_{j}}, X_{i_{l}}$ are descendants of the same node from depth-1 (i.e., the same depth-1 node is reached from $X_{i_{j}}, X_{i_{l}}$ by proceeding from child to parent). For example, the leaf nodes $X_{i_5}$ and $X_{i_6}$ are descendants of the same depth-1 node with label $X_{i_1}$. As $\{X_{i_1}, X_{i_5}\}$ can be used to decode $W_3$, we apply the same interference property to obtain that $X_{i_1}, X_{i_5}$ contain the same information about $W_2$, i.e.,
\begin{eqnarray}
\{X_{i_1}, X_{i_5}\} \in \mathcal{S}_3 \overset{\mbox{\scriptsize Property 2.(a)}}{\Longrightarrow}~ X_{i_1} \overset{W_2}{\simeq} X_{i_5}.
\end{eqnarray}
Similarly, $\{X_{i_3}, X_{i_6}\}$ can be used to decode $W_3$ so that they contain the same information about $W_2$,
\begin{eqnarray}
\{X_{i_6}, X_{i_3}\} \in \mathcal{S}_3 \overset{\mbox{\scriptsize Property 2.(a)}}{\Longrightarrow}~ X_{i_6} \overset{W_2}{\simeq} X_{i_3}.
\end{eqnarray}
Now suppose $X_{i_5} = X_{i_6}$. Applying the transitivity of the same information operation, we have that $X_{i_1}, X_{i_3}$ must contain the same information about $W_2$.
\begin{eqnarray}
X_{i_1} \overset{W_2}{\simeq} X_{i_5}, X_{i_5} \overset{W_2}{\simeq} X_{i_3} \overset{\mbox{\scriptsize Lemma \ref{lemma:trans}}}{\Longrightarrow} X_{i_1} \overset{W_2}{\simeq} X_{i_3}.
\end{eqnarray}
However, $\{X_{i_1}, X_{i_3}\}$ can be used to decode $W_2$, so from the distinct desired information property (Property 2.(b)), they must contain distinct information about $W_2$.
\begin{eqnarray}
\{X_{i_1}, X_{i_3}\} \in \mathcal{S}_2 \overset{\mbox{\scriptsize Property 2.(b)}}{\Longrightarrow}~ X_{i_1} \overset{W_2}{\perp} X_{i_3}.
\end{eqnarray}
Finally, we arrive at the contradiction by invoking the incompatibility property of same and distinct information (Property 3).
\begin{eqnarray}
X_{i_1} \overset{W_2}{\simeq} X_{i_3}, X_{i_1} \overset{W_2}{\perp} X_{i_3} \overset{\mbox{\scriptsize Property 3}}{\Longrightarrow} \mbox{Contradiction.}
\end{eqnarray}
Therefore we conclude that $X_{i_5}$ and $X_{i_6}$ must be distinct. The proof for other choices of $X_{i_{j}}, X_{i_{l}}$ is similar. 
\item $X_{i_{j}}, X_{i_{l}}$ are descendants of the same node from depth-0. For example, the leaf nodes $X_{i_6}$ and $X_{i_8}$ are descendants of the same depth-0 node with label $X_{i_1}$. The remaining proof is similar to the one above, where we trace $X_{i_6}$ to $X_{i_1}$ (and $X_{i_8}$ to $X_{i_2}$) using decoding constraints of $W_2, W_3$ and argue that they must contain the same information about $W_1$. Then if $X_{i_6} = X_{i_8}$, $X_{i_1}$ and $X_{i_2}$ must contain the same information about $W_1$, contradicting the fact that they must contain distinct information about $W_1$ (as $X_{i_1}$ and $X_{i_2}$ form a decoding set of $W_1$). 
\begin{eqnarray}
 \{X_{i_6}, X_{i_3}\} \in \mathcal{S}_3 &\overset{\mbox{\scriptsize Property 2.(a)}}{\Longrightarrow}& X_{i_6} \overset{W_1}{\simeq} X_{i_3} \\
 \{X_{i_3}, X_{i_1}\} \in \mathcal{S}_2 &\overset{\mbox{\scriptsize Property 2.(a)}}{\Longrightarrow}& X_{i_3} \overset{W_1}{\simeq} X_{i_1} \\
&\overset{\mbox{\scriptsize Lemma \ref{lemma:trans}}}{\Longrightarrow}& X_{i_6} \overset{W_1}{\simeq} X_{i_1} \\
&\mbox{\small (Symmetrically)}& X_{i_8} \overset{W_1}{\simeq} X_{i_3} \\
 \mbox{\small Suppose}~X_{i_6} = X_{i_8} &\overset{\mbox{\scriptsize Lemma \ref{lemma:trans}}}{\Longrightarrow}& X_{i_1} \overset{W_1}{\simeq} X_{i_3}\\
 \{X_{i_1}, X_{i_3}\} \in \mathcal{S}_1 &\overset{\mbox{\scriptsize Property 2.(b)}}{\Longrightarrow}& X_{i_1} \overset{W_1}{\perp} X_{i_3} \\
 &\overset{\mbox{\scriptsize Property 3}}{\Longrightarrow}& \mbox{Contradiction.}
\end{eqnarray}
The proof for other choices of $X_{i_{j}}, X_{i_{l}}$ is similar. 
\end{enumerate}

The proof for the 3 cases is now complete. To sum up, any two coded symbols can not be the same, i.e., all $N^K = 8$ coded symbols are all distinct, so the code length for any capacity achieving ULDC must satisfy $M \geq N^K = 8$. The converse proof with $N=2, K=3$ is thus complete.

\subsection{General Proof for Arbitrary $N, K$}
The general proof for arbitrary $N,K$ is a simple generalization of that presented in the previous section. Consider a full $N$-ary tree with depth $K$ (refer to Figure \ref{fig:decodetree}), root node label $X_{i_1}$ and permutation $\pi$. There are $N^K$ coded symbols that appear as the leaf nodes. We show that they are all distinct.

To set up the proof by contradiction, let us assume there exist two coded symbols $X_j, X_{j'}$ such that $X_j = X_{j'}$. We have two cases.
\begin{enumerate}
\item $X_{j}, X_{j'}$ are siblings. In this case if $X_{j} = X_{j'}$, then $H(X_{j}|X_{j'}) = 0$. However, as $X_{j}, X_{j'}$ are siblings, they belong a decoding set of $W_{\pi_K}$. Applying the independence property of coded symbols (Property 2.(c)), we have $H(X_j) = H(X_{j}|X_{j'}) = 0$, which contradicts  the fact that $H(X_j) = L_x \neq 0$ (as the code is capacity achieving). Therefore $X_{j}, X_{j'}$ must be distinct.

\item $X_{j}, X_{j'}$ are descendants of the same node (denoted as $X_{j*}$) from depth-$k, k \in [0:K-2]$. We find the path from $X_j$ to $X_{j*}$ (by moving from chid to parent recursively). The path passes $K- k+1$ nodes (one each from depth-$k', k' \in [k: K]$).
\begin{eqnarray}
X_{j} - X_{j_1} - X_{j_2} \cdots - X_{j_l} - X_{\widetilde{j}} - \cdots - X_{\widetilde{j}} - X_{j*}.
\end{eqnarray}

Note that due to the construction of the full $N$-ary tree, the coded symbol in the parent node is always equal to the coded symbol in the leftmost child node. The nodes that appear in the path are initially distinct but after some steps, the node (nodes) that appear in the path will be equal to $X_{\widetilde{j}}$ (which might be the same as $X_{j*}$ if $X_{\widetilde{j}}$ is the leftmost child of $X_{j*}$). Any two distinct adjacent nodes in the path belong to a decoding set of some source symbol $W_{k'}, k' \in [k+2:K]$. Applying the same interference property to each such pair of nodes, we have
\begin{eqnarray}
 X_{j} \overset{W_{k+1}}{\simeq} X_{j_1},  X_{j_1} \overset{W_{k+1}}{\simeq} X_{j_2}, \cdots,  X_{j_l} \overset{W_{k+1}}{\simeq} X_{\widetilde{j}}  \overset{\mbox{\scriptsize Lemma \ref{lemma:trans}}}{\Longrightarrow} X_{j} \overset{W_{k+1}}{\simeq} X_{\widetilde{j}}
\end{eqnarray}

Symmetrically, we consider the path from $X_{j'}$ to $X_{j*}$,
\begin{eqnarray}
X_{j'} - X_{j'_1} - X_{j'_2} \cdots - X_{j'_{l'}} - X_{\widetilde{j'}} - \cdots - X_{\widetilde{j'}} - X_{j*}.
\end{eqnarray}
Similarly, we apply the same interference property to distinct adjacent nodes in the path as they belong to a decoding set of some source symbol other than $W_{k+1}$.
\begin{eqnarray}
 X_{j'} \overset{W_{k+1}}{\simeq} X_{j'_1},  X_{j'_1} \overset{W_{k+1}}{\simeq} X_{j'_2}, \cdots,  X_{j'_{l'}} \overset{W_{k+1}}{\simeq} X_{\widetilde{j'}}  \overset{\mbox{\scriptsize Lemma \ref{lemma:trans}}}{\Longrightarrow} X_{j'} \overset{W_{k+1}}{\simeq} X_{\widetilde{j'}}
\end{eqnarray}
Now if $X_j = X_{j'}$, then 
\begin{eqnarray}
X_{j} \overset{W_{k+1}}{\simeq} X_{\widetilde{j}}, X_{j} \overset{W_{k+1}}{\simeq} X_{\widetilde{j'}} \overset{\mbox{\scriptsize Lemma \ref{lemma:trans}}}{\Longrightarrow} X_{\widetilde{j}} \overset{W_{k+1}}{\simeq} X_{\widetilde{j'}}
\end{eqnarray}
However, this contradicts  the fact that $X_{\widetilde{j}}, X_{\widetilde{j'}}$ belong to a decoding set of the source symbol $W_{k+1}$ (as the two paths overlap at node $X_{j*}$).
\begin{eqnarray}
X_{\widetilde{j}} \overset{W_{k+1}}{\perp} X_{\widetilde{j'}}, X_{\widetilde{j}} \overset{W_{k+1}}{\simeq} X_{\widetilde{j'}} \overset{\mbox{\scriptsize Property 3}}{\Longrightarrow} \mbox{Contradiction.}
\end{eqnarray}
\end{enumerate}
Therefore, $X_j = X_j*$ can not hold and we have $N^K$ distinct coded symbols, i.e., $M \geq N^K$. The proof is thus complete. $\hfill\QED$

{\it Remark: Comparing our minimum length proof of capacity achieving ULDC (and the upload cost proof of PIR$_{\max}$) to the upload cost proof of PIR$_{\mbox{\footnotesize ave}}$ \cite{Tian_Sun_Chen_Upload}, we have an additional non-zero entropy property (Property 1 in Lemma \ref{lemma:cap}) that allows the proof to work for all non-linear schemes (whereas the result of \cite{Tian_Sun_Chen_Upload} is limited to a restricted class of decomposable schemes).} 

\subsection{Proof of Lemma \ref{lemma:cap}}\label{sec:lemmacap}
Let us prove the properties listed in Lemma \ref{lemma:cap} one at a time.
\subsubsection{Proof of Property $1$}
To set up the proof by contradiction, let us  assume, for some $i_1\in[1:M], k\in[1:K]$,
\begin{align}
H(X_{i_1} | W_{\overline{k}}) = 0.\label{eq:assume}
\end{align}

Consider a full $N$-ary tree (see Figure \ref{fig:decodetree}) with root node label $X_{i_1}$ and permutation $\pi$ such that $\pi_1 = k$. Thus $W_{\pi_{2:K}}=W_{\bar{k}}$. For a capacity achieving ULDC, all the inequalities from \eqref{eq:p1} to \eqref{eq:pf} must be equalities. Replacing \eqref{eq:p2} and \eqref{eq:pf} with equalities, we have
\begin{align}
L_w&=\sum_{X_i \in {S}_{\pi_1}^{[1]}} H(X_{i} | W_{\overline{k}}) \\
&= H(X_{i_1} | W_{\overline{k}})+H(X_{i_2} | W_{\overline{k}}) + \cdots + H(X_{i_{N-1}} | W_{\overline{k}})+H(X_{i_{N}} | W_{\overline{k}})\\
&=H(X_{i_2} | W_{\overline{k}}) + \cdots +H(X_{i_{N}} | W_{\overline{k}})\label{eq:term}
\end{align}
where in (\ref{eq:term}), we used our assumption \eqref{eq:assume}.
Because the sum of $N-1$ non-negative terms is equal to $L_w$, we must have at least one term, say corresponding to $X_{i*}$, that is not less than $\frac{L_w}{N-1}$. 
\begin{eqnarray}
H(X_{i*} | W_{\overline{k}}) \geq \frac{L_w}{N-1}. \label{eq:c1}
\end{eqnarray}
Because the code is universal, there exists a decoding set $S \in \mathcal{S}_j$ of message $W_j, j\neq k$ that contains $X_{i*}$.
\begin{eqnarray}
NL_x &=& \sum_{X_i \in S} H(X_i) \\
&\overset{(\ref{eq:lemmadec})}{\geq}& L_w + H(X_{i^*} | W_j) \\
&\overset{}{\geq}& L_w + H(X_{i^*} | W_{\overline{k}}) 
\end{eqnarray}
Plugging in the capacity achieving condition, $L_x =  C_{\mbox{\tiny ULDC}}^*(N,K)^{-1} L_w$, we have
\begin{eqnarray}
H(X_{i^*} | W_{\overline{k}})  &\leq& L_w( N  C_{\mbox{\tiny ULDC}}^*(N,K)^{-1} -1) \\
&=& \left(\frac{1}{N} + \frac{1}{N^2} + \cdots + \frac{1}{N^{K-1}}\right)L_w \\
&<& \frac{1/N}{1-1/N} L_w = \frac{L_w}{N-1} \label{eq:c2}
\end{eqnarray}
But  (\ref{eq:c1}) and (\ref{eq:c2}) contradict each other. The contradiction completes the proof of Property 1. $\hfill\QED$

\subsubsection{Proof of Property 2}
First let us prove $(a)$,  that $\forall X_{i_1}, X_{i_2}\in S\in \mathcal{S}_k$ and $\forall k'\neq k$, $X_{i_{1}} \overset{W_{k'}}{\simeq} X_{i_2}$. For this purpose, let us consider a full $N$-ary tree (see Figure \ref{fig:decodetree}) where the root has label $X_{i_1}$,  the permutation $\pi$ satisfies $\pi_K = k$, and $X_{i_1}, X_{i_2}$ appear at depth-$K$ in decoding set $S$. Consider the step from depth-$K$ to depth-$(K-1)$ of the converse proof (i.e., (\ref{eq:p1})). As we assume the ULDC achieves the capacity, the following equality must hold (refer to (\ref{eq:l1})).
\begin{align}
\sum_{X_i\in S_{\pi_K}}H(X_i)&=\sum_{X_i\in S}H(X_i)\\
&= L_w+H(X_{i_1}\mid W_k) \label{eq:t0}\\
&= L_w+H(S\mid W_k)\label{eq:t1}
\end{align}
In \eqref{eq:t1} we used \eqref{eq:l1}, which must also be an equality for a capacity achieving ULDC. From \eqref{eq:t0} and \eqref{eq:t1} we must have
\begin{eqnarray}
&& H(X_{i_1}, X_{i_2} | W_k ) = H(X_{i_1} | W_k) \\
&\Rightarrow&  H(X_{i_2} | X_{i_1}, W_{k}) = 0 \\
&\Rightarrow& H(X_{i_2} | X_{i_1}, W_{\overline{k'}}) = 0, k' \neq k.
\end{eqnarray}
By symmetry, we can similarly prove $H(X_{i_1} | X_{i_2}, W_{\overline{k'}}) = 0$ so that $X_{i_{1}} \overset{W_{k'}}{\simeq} X_{i_2}$ and we have proved Property $2(a)$.

To prove Property $2(b)$, we consider a full $N$-ary tree (see Figure \ref{fig:decodetree}) where the root has label $X_{i_1}$, the permutation $\pi$ satisfies $\pi_1 = k$ (such that $\pi_{2:K} = \overline{k}$), and the label $X_{i_2}$ appears at depth-$1$. 
Consider the step from depth-$1$ to depth-$0$ of the converse proof (i.e., (\ref{eq:p2})). As the ULDC achieves the capacity, the following equality must hold (refer to (\ref{eq:l2})).
\begin{eqnarray}
&& H(X_{i_1} | W_{\overline{k}}) + H(X_{i_2} | W_{\overline{k}}) = H(X_{i_1}, X_{i_2} | W_{\overline{k}}) \\
&\Rightarrow& H(X_{i_1} | X_{i_2}, W_{\overline{k}}) = H(X_{i_1} | W_{\overline{k}})
\end{eqnarray}
Therefore we have proved Property $2(b)$, that $X_{i_{1}} \overset{W_{k}}{\perp} X_{i_2}$ holds.

To prove Property $2(c)$, we consider a full $N$-ary tree (see Figure \ref{fig:decodetree}) where the root label is $X_{i_1}$,  the permutation $\pi$ satisfies $\pi_K = k$, and the label $X_{i_2}$ appears at depth-$K$. %The term $H(X_i | W_{\overline{k}}) = H(X_i | W_{\pi_{2:K}})$ appears (as the first node) in depth-1. 
Consider the step from depth-$K$ to depth-$(K-1)$ of the converse proof (i.e., (\ref{eq:p1})). As we assume the ULDC achieves the capacity, the following equality must hold (refer to (\ref{eq:l2})).
\begin{eqnarray}
&& H(X_{i_1} ) + H(X_{i_2} ) = H(X_{i_1}, X_{i_2} ) \\
&\Rightarrow& H(X_{i_1} | X_{i_2}) = H(X_{i_1})
\end{eqnarray}
Therefore the desired claim is proved. $\hfill\QED$

\subsubsection{Proof of Property $3$}
\begin{eqnarray}
&& X_{i_{1}} \overset{W_{k}}{\simeq} X_{i_2} ~\Rightarrow~ H(X_{i_1} | X_{i_2}, W_{\overline{k}}) = 0 \\
&& X_{i_{1}} \overset{W_{k}}{\perp} X_{i_2}  ~\Rightarrow~ H(X_{i_1} | X_{i_2}, W_{\overline{k}}) = H(X_{i_1} | W_{\overline{k}}) \\
&\Rightarrow& H(X_{i_1} | W_{\overline{k}}) = 0
\end{eqnarray}
which contradicts the non-zero entropy property (Property 1). So same and distinct information conditions can not be simultaneously satisfied and the proof is complete.$\hfill\QED$

\section{Proof of Theorem \ref{thm:length}: Achievability}\label{sec:length_ach}
In this section, we present the construction of a capacity achieving SLDC with length $M = N^K$. Before proceeding to the general proof, we first consider two examples.
\subsection{Example 1: $N=2, K=2$}
When $N=2, K =2$, the capacity is $C_{\mbox{\tiny ULDC}}^*(N=2,K=2) = \frac{L_w}{L_x} = 2(1+\frac{1}{2})^{-1} = \frac{4}{3}$. We present an SLDC with length $4$, where each source symbol is comprised of $L_w = 4$ bits and each coded symbol has $L_x = 3$ bits.

Denote $W_1 = (a_1, a_2, a_3, a_4), W_2 = (b_1, b_2, b_3, b_4)$, where $a_i, b_j$ are i.i.d. uniform bits. The code is as follows.
\begin{eqnarray}
\begin{array}{c | c | c | c}
X_1 & X_2 & X_3 & X_4 \\ \hline
\emptyset & a_1 & a_1 + b_1 & b_1 \\
a_2 & \emptyset & b_2 & a_2 + b_2 \\
b_3 & a_3 + b_3 & a_3 & \emptyset \\
a_4 + b_4 & b_4 & \emptyset & a_4 
\end{array} \label{eq:c22}
\end{eqnarray}

We have $2$ decoding sets for each source symbol.
\begin{eqnarray}
{\mathcal{S}_1} = \{ \{X_1, X_2\}, \{X_3, X_4\} \} \\
{\mathcal{S}_2} = \{ \{X_1, X_4\}, \{X_2, X_3\} \}
\end{eqnarray}
Correctness is easy to verify (i.e., from any decoding in ${\mathcal{S}_k}$, we can decode $W_k$). Perfect smoothness is also easily verified, as each coded symbol appears once and only once in the decoding sets for any message.

Inspecting the code in (\ref{eq:c22}), we see that each row forms a feasible sub-code and the rows are some permutations of each other (note however, this is a highly-structured permutation that preserves the same upload cost and is particularly distinct from time-sharing). This is in fact the key idea of our SLDC and we will further develop it in the following example and in the general proof.

\subsection{Example 2: $N=3, K=3$}
When $N=3, K=3$, the capacity is $C_{\mbox{\tiny ULDC}}^*(N=3,K=3) = \frac{L_w}{L_x} = 3(1+\frac{1}{3} + \frac{1}{3^2})^{-1} = \frac{27}{13} = \frac{54}{26}$. We present an SLDC with length $27$, where each source symbol is comprised of $L_w = 54$ bits and each coded symbol has $L_x = 26$ bits.

Each source symbol is divided into $27$ sub-source-symbols and each sub-source-symbol has $2$ bits. Denote $W_1$ as the collection of $(a_1^{(\gamma_1, \gamma_2, \gamma_3)}, a_2^{(\gamma_1, \gamma_2, \gamma_3)})$ for all $\gamma_1, \gamma_2, \gamma_3$, where $\gamma_1, \gamma_2, \gamma_3 \in [0:2]$ are indices for sub-source-symbol. Similarly, $W_2$ is the collection of $(b_1^{(\gamma_1, \gamma_2, \gamma_3)}, b_2^{(\gamma_1, \gamma_2, \gamma_3)})$ and $W_3$ is the collection of $(c_1^{(\gamma_1, \gamma_2, \gamma_3)}, c_2^{(\gamma_1, \gamma_2, \gamma_3)})$. $a_i, b_j, c_l$ are i.i.d. uniform bits.
$a_0^{(\gamma_1, \gamma_2, \gamma_3)}, b_0^{(\gamma_1, \gamma_2, \gamma_3)}, c_0^{(\gamma_1, \gamma_2, \gamma_3)}$ are set to 0.

To simplify the notation, we denote the $N^K = 27$ coded symbols as $X_{p_1, p_2, p_3}$ where $p_i \in [0:2], i \in [1:3]$. {\color{black} These 27 coded symbols are divided into $3$ groups depending on the value of $p_1+p_2+p_3$, so that $x_{p_1,p_2,p_3}$ belongs to Group $p_1 + p_2 + p_3$ (modulo 3), and each group has $9$ coded symbols. }

Each coded symbol is similarly comprised of $27$ sub-coded-symbols, denoted as $X_{p_1, p_2, p_3}^{(\gamma_1, \gamma_2, \gamma_3)}$. When there will be no confusion from the context, we simply denote $X_{p_1, p_2, p_3}^{(\gamma_1, \gamma_2, \gamma_3)}$ as $x_{p_1,p_2,p_3}$. To determine the value of $x_{p_1,p_2,p_3}$, we use $p_k + \gamma_k$ as the bit sub-script for the $(\gamma_1, \gamma_2, \gamma_3)$ sub-source-symbol of $W_k, k \in [1:3]$ and take the sum of all $3$ bits, i.e., $x_{p_1,p_2,p_3} = a_{p_1 + \gamma_1}^{(\gamma_1, \gamma_2, \gamma_3)} + b_{p_2 + \gamma_2}^{(\gamma_1, \gamma_2, \gamma_3)} + c_{p_3 + \gamma_3}^{(\gamma_1, \gamma_2, \gamma_3)}$. {\color{black} For example, the symbol denoted as $x_{0,1,2}=a_{\gamma_1}+b_{1+\gamma_2}+c_{2+\gamma_3}$, is comprised of $27$ sub-coded-symbols corresponding to all $27$ values of $(\gamma_1, \gamma_2,\gamma_3)\in[0:2]^3$, such as $a_1+b_0+c_1$ when $(\gamma_1,\gamma_2,\gamma_3)=(1,2,2)$. All these symbols belong to Group $0$ because $p_1+p_2+p_3=0+1+2=0$ mod $3$.}

\begin{eqnarray}
\begin{array}{l | l | l }
\mbox{\small Group 0} & \mbox{\small Group 1} & \mbox{\small Group 2} \\ \hline
x_{0,0,0} = a_{\gamma_1} + b_{\gamma_2} + c_{\gamma_3}& x_{0,0,1} = a_{\gamma_1} + b_{\gamma_2} + c_{1+\gamma_3} & x_{0,0,2} = a_{\gamma_1} + b_{\gamma_2} + c_{2+\gamma_3} \\
x_{1,1,1} = a_{1+\gamma_1} + b_{1+\gamma_2} + c_{1+\gamma_3} & x_{0,1,0} = a_{\gamma_1} + b_{1+\gamma_2} + c_{\gamma_3} & x_{0,2,0} = a_{\gamma_1} + b_{2+\gamma_2} + c_{\gamma_3} \\
x_{2,2,2} = a_{2+\gamma_1} + b_{2+\gamma_2} + c_{2+\gamma_3} & x_{1,0,0} = a_{1+\gamma_1} + b_{\gamma_2} + c_{\gamma_3} & x_{2,0,0} = a_{2+\gamma_1} + b_{\gamma_2} + c_{\gamma_3} \\
x_{0,1,2} = a_{\gamma_1} + b_{1+\gamma_2} + c_{2+\gamma_3} & x_{0,2,2} = a_{\gamma_1} + b_{2+\gamma_2} + c_{2+\gamma_3} & x_{0,1,1} = a_{\gamma_1} + b_{1+\gamma_2} + c_{1+\gamma_3} \\
x_{0,2,1} = a_{\gamma_1} + b_{2+\gamma_2} + c_{1+\gamma_3} & x_{2,0,2} = a_{2+\gamma_1} + b_{\gamma_2} + c_{2+\gamma_3} & x_{1,0,1} = a_{1+\gamma_1} + b_{\gamma_2} + c_{1+\gamma_3} \\
x_{1,0,2} = a_{1+\gamma_1} + b_{\gamma_2} + c_{2+\gamma_3} & x_{2,2,0} = a_{2+\gamma_1} + b_{2+\gamma_2} + c_{\gamma_3} & x_{1,1,0} = a_{1+\gamma_1} + b_{1+\gamma_2} + c_{\gamma_3} \\
x_{2,0,1} = a_{2+\gamma_1} + b_{\gamma_2} + c_{1+\gamma_3} & x_{1,1,2} = a_{1+\gamma_1} + b_{1+\gamma_2} + c_{2+\gamma_3} & x_{2,2,1} = a_{2+\gamma_1} + b_{2+\gamma_2} + c_{1+\gamma_3} \\
x_{1,2,0} = a_{1+\gamma_1} + b_{2+\gamma_2} + c_{\gamma_3} & x_{1,2,1} = a_{1+\gamma_1} + b_{2+\gamma_2} + c_{1+\gamma_3} & x_{2,1,2} = a_{2+\gamma_1} + b_{1+\gamma_2} + c_{2+\gamma_3} \\
x_{2,1,0} = a_{2+\gamma_1} + b_{1+\gamma_2} + c_{\gamma_3} & x_{2,1,1} = a_{2+\gamma_1} + b_{1+\gamma_2} + c_{1+\gamma_3} & x_{1,2,2} = a_{1+\gamma_1} + b_{2+\gamma_2} + c_{2+\gamma_3} 
\end{array}
\end{eqnarray}

The decoding constraints are as follows (easy to verify from the  table above).
\begin{eqnarray}
&& \mbox{From}~x_{p_1, p_2, p_3}, x_{p_1+1, p_2, p_3}, x_{p_1+2,p_2,p_3},~\mbox{we can decode}~a_{p_1}, a_{p_1+1}, a_{p_1+2}. \\
&& \mbox{From}~x_{p_1, p_2, p_3}, x_{p_1, p_2+1, p_3}, x_{p_1,p_2+2,p_3},~\mbox{we can decode}~b_{p_2}, b_{p_2+1}, b_{p_2+2}. \\
&& \mbox{From}~x_{p_1, p_2, p_3}, x_{p_1, p_2, p_3+1}, x_{p_1,p_2,p_3+2},~\mbox{we can decode}~c_{p_3}, c_{p_3+1}, c_{p_3+2}. 
\end{eqnarray}
That is, if we pick one coded symbol from each group such that their subscripts only differ in the $k^{th}$ digit, then we can decode $W_k$. Further, this claim remains valid for any realization of $(\gamma_1, \gamma_2, \gamma_3)$.
As a result, for each source symbol, we have $9$ decoding sets and each coded symbol appears once and only once in the decoding sets, leading to correctness and perfect smoothness. 

Finally, we note that each coded symbol contains 26 bits, although it contains 27 sub-coded-symbols (each sub-coded-symbol is one equation, thus at most 1 bit). This follows from the observation that for any $p_1, p_2, p_3$, there exists one and only one realization of $(\gamma_1, \gamma_2, \gamma_3)$ such that $p_i+\gamma_i = 0~\mbox{(modulo 3)}, \forall i \in [1:3]$, $X_{p_1, p_2, p_3}^{(\gamma_1, \gamma_2, \gamma_3)} = a_0 + b_0 + c_0 = 0$ and nothing needs to be stored. For all other cases, the sub-coded-symbol is 1 bit. Therefore, $L_x = 26$ and the SLDC achieves the capacity.

\subsection{General Proof for Arbitrary $N,K$}
The general proof follows from the ideas presented in previous sections.
For any $N,K$, the capacity is $C_{\mbox{\tiny ULDC}}^*(N,K) = \frac{L_w}{L_x} = N(1+\frac{1}{N}+\cdots+\frac{1}{N^{K-1}})^{-1} = \frac{N^K(N-1)}{N^K-1}$. We present an SLDC with length $M = N^K$, where each source symbol is comprised of $L_w = N^K(N-1)$ bits and each coded symbol has $L_x = N^K-1$ bits.

Each source symbol is divided into $N^K$ sub-source-symbols and each sub-source-symbol has $N-1$ bits. Define $\vec{\gamma} = (\gamma_1, \gamma_2, \cdots, \gamma_{K})$.
\begin{eqnarray}
W_k &=& (W_{k}^{(0,0,\cdots,0)}, W_{k}^{(0,0,\cdots,1)}, \cdots, W_{k}^{(N-1,N-1,\cdots,N-1)}), \forall k \in [1:K]\\
W_k^{\vec{\gamma}} &=& (W_{k,0}^{\vec{\gamma}}, W_{k,1}^{\vec{\gamma}}, W_{k,2}^{\vec{\gamma}}, \cdots, W_{k,N-1}^{\vec{\gamma}}), \forall i \in [1:K], \forall \gamma_i \in [0:N-1]\\
W_{k,0}^{\vec{\gamma}} &\triangleq& 0 \label{eq:w0}
\end{eqnarray}

Define $\vec{p} = (p_1, p_2, \cdots, p_K)$.
The $N^K$ coded symbols are denoted as $X_{\vec{p}}$, where $i \in [1:K], p_i \in [0:N-1]$. These $N^K$ coded symbols are divided into $N$ groups depending on the value of $\sum_{i=1}^K p_i$ (modulo $N$), so that $X_{\vec{p}}$ belongs to Group $\sum_{i=1}^K p_i ~\mbox{(modulo $N$)}$ and each group has $N^{K-1}$ coded symbols.
\begin{eqnarray}
\forall n \in [0:N-1], ~\mbox{Group}~n = \Bigg\{X_{\vec{p}}: \sum_{i=1}^K p_i ~\mbox{(modulo $N$)}~ = n \Bigg\}. \label{eq:groupn}
\end{eqnarray}

Each coded symbol is similarly comprised of $N^K$ sub-coded-symbols and each sub-coded-symbol is designed as follows.
\begin{eqnarray}
X_{\vec{p}} &=& (X_{\vec{p}}^{(0,0,\cdots,0)}, X_{\vec{p}}^{(0,0,\cdots,1)}, \cdots, X_{\vec{p}}^{(N-1,N-1,\cdots,N-1)}), \forall \vec{p} \\
X_{\vec{p}}^{\vec{\gamma}} &=& W_{1, p_1+\gamma_1}^{\vec{\gamma}} + W_{2, p_2+\gamma_2}^{\vec{\gamma}} + \cdots + W_{K, p_K+\gamma_K}^{\vec{\gamma}}, \forall \vec{\gamma} \label{eq:xk}
\end{eqnarray}

For each message, we have $N^{K-1}$ decoding sets. For given $p_1, \cdots, p_{k-1}, p_{k+1}, \cdots, p_{K}$, define $p^*_k = N - (p_1 + \cdots+ p_{k-1} + p_{k+1} + \cdots + p_{K})$ (modulo $N$). The subscripts below are understood modulo $N$.
\begin{eqnarray}
&&\forall k \in [1:K], \forall i \in [1:k-1] \cup [k+1:K], \forall p_i \in [0:N-1], \\
&& \mathcal{S}_k = \bigcup_{\forall p_i, i\neq k} \big\{X_{p_{1}, \cdots, p_{k-1}, p^*_k, p_{k+1}, \cdots, p_K}, X_{p_{1}, \cdots, p_{k-1}, p^*_k+1, p_{k+1}, \cdots, p_K}, \notag\\
&&~~~~~~~~~~~~~~~~~\cdots, X_{p_{1}, \cdots, p_{k-1}, p^*_k+N-1, p_{k+1}, \cdots, p_K} \big\} \label{eq:xs}
\end{eqnarray}
where each decoding set is comprised of one and only one coded symbol from each group.

We verify that the code is correct, perfectly smooth and capacity achieving.

First, to show that the code is correct, we verify that from any coding set in $\mathcal{S}_k$, we can decode $W_k, \forall k \in [1:K]$. Consider any realization of $p_1, \cdots, p_{k-1}, p_{k+1}, \cdots, p_{K}$. From (\ref{eq:xk}), we consider the $N$ coded symbols and obtain that $\forall \vec{\gamma}$,
\begin{eqnarray}
&& X_{p_{1}, \cdots, p_{k-1}, p^*_k, p_{k+1}, \cdots, p_K}^{\vec{\gamma}} =  \sum_{j=1, j \neq k}^K W_{j, p_j + \gamma_j}^{\vec{\gamma}} + W_{k, p_k^* + \gamma_k}^{\vec{\gamma}} \\
&& X_{p_{1}, \cdots, p_{k-1}, p^*_k+1, p_{k+1}, \cdots, p_K}^{\vec{\gamma}} = \sum_{j=1, j \neq k}^K W_{j, p_j + \gamma_j}^{\vec{\gamma}} + W_{k, p_k^* +1 + \gamma_k}^{\vec{\gamma}} \\ 
&& \cdots \\
&& X_{p_{1}, \cdots, p_{k-1}, p^*_k+N-1, p_{k+1}, \cdots, p_K}^{\vec{\gamma}} = \sum_{j=1, j \neq k}^K W_{j, p_j + \gamma_j}^{\vec{\gamma}} + W_{k, p_k^*+ N-1 + \gamma_k}^{\vec{\gamma}}
\end{eqnarray}
Note that  the interference about source symbols $W_{\overline{k}}$ is the same in the above $N$ equations and the desired sub-source-symbol has $N-1$ bits. So we can decode all $N-1$ desired bits, $W_{k, 1}^{\vec{\gamma}}, W_{k, 2}^{\vec{\gamma}}, \cdots, W_{k, N-1}^{\vec{\gamma}}$. Repeating the same decoding procedure for all $\vec{\gamma}$, we  decode all $L_w = N^K(N-1)$ bits in $W_k$. Therefore the LDC is correct.

Second, the code is perfectly smooth because from (\ref{eq:xs}), we note that for any source symbol $W_k$ and for any Group $n \in [0:N-1]$, any coded symbol $X_{\vec{p}}$ (from Group $n$) appears once and only once. Therefore, the definition of perfect smoothness (refer to Definition \ref{def:4}) is satisfied.

Finally, we prove that the code achieves the capacity. To this end, we verify that $H(X_{\vec{p}}) = L_x = N^K-1, \forall \vec{p}$. Note that each coded symbol contains $N^K$ sub-coded-symbols, and there exists one and only one sub-coded-symbol that is constantly zero. That is, for any given $\vec{p}$, when
\begin{eqnarray}
\gamma_k = - p_k ~\mbox{(modulo $N$)}, \forall k \in [1:K],
\end{eqnarray}
we have $X_{\vec{p}}^{\vec{\gamma}} = \sum_{k=1}^K W_{k, 0}^{\vec{\gamma}} = 0$ (refer to (\ref{eq:xk}), (\ref{eq:w0})). The proof is thus complete. $\hfill\QED$

{\it Remark: One might wonder if our SLDC (and the corresponding upload optimal PIR$_{\max}$ scheme) can be constructed from the upload optimal PIR$_{\mbox{\footnotesize ave}}$ scheme in \cite{Tian_Sun_Chen_Upload} by symmetrization (e.g., as described in Section 5 of \cite{Tian_Sun_Chen_Upload}), as one sub-code in our scheme is similar to the PIR$_{\mbox{\footnotesize ave}}$ scheme in \cite{Tian_Sun_Chen_Upload}. This does not work because general symmetrization techniques will increase the upload proportional to the number of concatenations of sub-codes, while in our PIR$_{\max}$ scheme, the upload cost of the concatenated code remains the same as that of one sub-code (i.e., $(K-1) \log (N)$ per database). Therefore, our code is is not constructed by generic symmetrizations. Instead, the specific sub-code has a permutation-invariant property that allows us to shift the symbol indices while retaining the same decoding structure (refer to (\ref{eq:xk})).}

\section{Proof of Corollaries \ref{cor:rir} and  \ref{cor:pir}}\label{sec:pir}
For the converse, it suffices to provide the proof for RIR$_{\max}$, which automatically implies the converse for PIR$_{\max}$. The converse proof for RIR$_{\max}$ is as follows.

To set up the proof by contradiction, suppose on the contrary that we have a capacity achieving RIR$_{\max}$ scheme such that the upload cost from some database is strictly less than $(K-1)\log(N)$, i.e., there exists a set of answers $\mathcal{X}^{[n]}$ from one database such that $|\mathcal{X}^{[n]} | < N^{K-1}$.
Then by Observation \ref{thm:LDCPIR}, we have a capacity achieving ULDC such that there exists at least one group of strictly fewer than $N^{K-1}$ coded symbols (this group corresponds to the set of answers $\mathcal{X}^{[n]}$ from the database in PIR) such that any decoding set must contain one coded symbol from this group (as any decoding set in PIR must contain one answer from each database, including the one with answer set $\mathcal{X}^{[n]}$).  Note that for any full $N$-ary tree (refer to Figure \ref{fig:decodetree}), the $N^K$ leaf nodes form $N^{K-1}$ decoding sets.
As any one of these $N^{K-1}$ decoding sets must contain one coded symbol from $\mathcal{X}^{[n]}$ (where $|\mathcal{X}^{[n]}|<N^{K-1}$), the leaf nodes must have at least two identical coded symbols. Then from the converse proof of Theorem \ref{thm:length}, it follows that the ULDC can not achieve capacity and we arrive at the contradiction.

For the achievability, it suffices to provide the proof for PIR$_{\max}$, which automatically implies the achievability for RIR$_{\max}$. The achievable scheme for PIR$_{\max}$ is  based on the SLDC from Theorem \ref{thm:length}. The SLDC has an $N$-partite property, that \emph{any decoding set is comprised of one symbol from each group.} Group $n, n\in[0:N-1]$ maps to answer set $\mathcal{X}^{[n+1]}$, i.e., the coded symbols from Group $n, n \in [0:N-1]$ of the SLDC (refer to (\ref{eq:groupn})) form the answers from the $(n+1)^{th}$ database in PIR$_{\max}$. The decoding supersets $\mathcal{S}_{[1:K]}$ of PIR$_{\max}$ are chosen to be the same as the decoding supersets $\mathcal{S}_{[1:K]}$ of the SLDC. Now if the user wishes to retrieve $W_k$, the user simply asks for one of the decoding sets for $W_k$ of the SLDC, uniformly over all $N^{K-1}$ choices of decoding sets (refer to (\ref{eq:xs})). Thus, the user downloads exactly  $N$ answers, one from each database. The correctness and perfect smoothness of LDC translate to the correctness and privacy of PIR$_{\max}$ directly. $\hfill\QED$

\section{Discussion}\label{sec:conc}
We introduce the notion of capacity for LDC, and show that the capacity of ULDCs and SLDCs with $K$ source symbols and locality $N$ is $C^* = N\left(1+1/N+1/N^2+\cdots+1/N^{K-1}\right)^{-1}$. We further show that the minimum length of capacity achieving ULDCs and SLDCs is $N^K$. The results are translated into the context of PIR$_{\max}$ and RIR$_{\max}$, where we show that the capacity of RIR$_{\max}$ is equal to that of PIR$_{\max}$, and the minimum upload cost of both PIR$_{\max}$ and RIR$_{\max}$ is equal to $(K-1)\log N$.

In this work, we have focused on the capacity achieving regime for LDCs. That is, the number of bits in each coded symbol is equal to $1/C^*$ times the number of bits in each source symbol, $L_x = \frac{L_w}{C^*} = \frac{L_w(1-1/N^K)}{N-1} < \frac{L_w}{N-1}$. In other words, the size of each coded symbol is (sometimes much) smaller than the size of each source symbol, a regime that is rarely studied in classical coding theory or theoretical computer science. Specifically, when the coded symbol has the smallest size (capacity achieving), the code length $M$ must be exponential, %in the locality $N$, 
i.e., $M \geq N^K$ in order to preserve either universality or perfect smoothness. It is an interesting avenue for future work to study other rate regimes. In particular, the minimum symbol rate for which the code length is polynomial %in the locality 
remains an interesting question.

As a final remark, we note that in the PIR$_{\max}$ problem formulation of this work, we have defined the $\max$ to be over all queries \emph{and} all databases, as this formulation is the one that connects to LDCs and is consistent with most scenarios. Essentially, we restrict the downloads to be symmetric and constant over all databases. An alternative formulation could be defining the $\max$ to be only over all queries, e.g., this formulation was adopted in \cite{Sun_Jafar_PIRL}, where the downloads are constant for one database, but could be asymmetric across the databases. These two formulations have the same capacity, but could behave differently in terms of other metrics, such as message size, upload cost etc. It is an interesting question to compare these models and identify their similarities and differences.

\bibliographystyle{IEEEtran}
\bibliography{Thesis}
\end{document}